%% file: ms.tex
\documentclass[a4paper]{article}
\pdfoutput=1
\usepackage[breaklinks=true,bookmarks=true,colorlinks=true,
            linkcolor=blue,citecolor=blue,allcolors=blue]{hyperref}
\usepackage{etex}
\usepackage{amsmath,amsbsy,amssymb,amsgen,amsfonts}
\usepackage[pdftex]{graphicx}

\usepackage{color}

\usepackage[left=1in, right=1in, top=1in,bottom=1in]{geometry}

\usepackage{natbib}

\newcommand{\revision}[2]{{#2}}
\newcommand{\revisionold}[2]{{#2}}
\input{lines_and_symbols}
\input{nomenclature}

\title{Formation of sediment patterns in channel flow:
    \revision{The minimal ripple unit and beyond }{%
      minimal unstable systems and their temporal evolution
    }}

\author{Aman G.\
  Kidanemariam\footnote{\href{mailto:aman.kidanemariam@kit.edu}{aman.kidanemariam@kit.edu}}
  \hspace*{1ex}and 
  Markus
  Uhlmann\footnote{\href{mailto:markus.uhlmann@kit.edu}{markus.uhlmann@kit.edu}} 
  \\[1ex]
  {\small 
    Institute for Hydromechanics, Karlsruhe Institute of
    Technology, 76131 Karlsruhe, Germany}\\
  {\small(Dated: \today\ -- accepted for publication in \textit{J.\ Fluid Mech})}
}
\date{}
\begin{document}

\maketitle

\input{abstract}
\input{introduction}
\input{numerics}
\input{setup}

\input{dune_dimensions}
\input{result}
\input{conclusion}
\input{acknowledgments}
\appendix
\input{appendix}

\bibliographystyle{jfm}
\addcontentsline{toc}{section}{References}
\bibliography{ms.bbl}
\end{document}

%% file: lines_and_symbols.tex
\usepackage{relsize}
\def\asolidthick{\protect\rule[2pt]{10.pt}{1pt}}

\newcommand{\aopencircle}
           {$\mathlarger{\mathlarger{\mathlarger{\circ}}}$}

\newcommand{\aopensquare}{$\mathsmaller{\square}$}
\newcommand{\asolidcircle}
           {$\mathlarger{\mathlarger{\mathlarger{\bullet}}}$}
\newcommand{\asolidtriup}{$\blacktriangle$}

\newcommand{\asolidtriright}{$\blacktriangleright$}
\newcommand{\asolidtrileft}{$\blacktriangleleft$}
      
\newcommand{\asolidsquare}{$\mathsmaller{\blacksquare}$}

\newcommand{\across}{$\mathlarger{\mathlarger{\times}}$}
\definecolor{amanred}{rgb}{0.8,0,0}
\definecolor{amanblue}{rgb}{0,0,0.7}
\definecolor{myblue}{rgb}{0,0.4,0.8}
\definecolor{myred}{rgb}{0.63,0.12,0.1}
\definecolor{mygreen}{rgb}{0,0.6,0.51}
\definecolor{mygray}{rgb}{0.63,0.63,0.63}
\definecolor{myorange}{rgb}{0.9,0.55,0.0}
\definecolor{myturk}{rgb}{0.13,0.42,0.52}
\definecolor{mypurple}{rgb}{0.66,0.42,0.61}

%% file: nomenclature.tex
\newcommand\caseA{H3}   
 \newcommand\caseB{H4}          
\newcommand\caseBa{H4\textsuperscript{1}}     
\newcommand\caseBb{H4\textsuperscript{2}} 
\newcommand\caseBc{H4\textsuperscript{3}} 
\newcommand\caseC{H6}         
\newcommand\caseD{H7}   
\newcommand\caseE{H12}
\newcommand\caseEa{H12\textsuperscript{1}}
\newcommand\caseEb{H12\textsuperscript{2}}
\newcommand\caseEc{H12\textsuperscript{3}}  
\newcommand\caseF{H48}  
\newcommand\Lx{\ensuremath{L_x}}
\newcommand\Ly{\ensuremath{L_y}}
\newcommand\Lz{\ensuremath{L_z}}
\newcommand\Npt{\ensuremath{N_p}}          %

\newcommand\xtilde{\ensuremath{\tilde{x}}}

\newcommand\utau{\ensuremath{u_\tau}}
\newcommand\Reb{\ensuremath{Re_b}}   %
\newcommand\Ret{\ensuremath{Re_\tau}}   %
\newcommand\Ga{\ensuremath{Ga}}    %
\newcommand\flowrate{\ensuremath{q_f}}    %
\newcommand\partflowrate{\ensuremath{\langle q_p\rangle}}    %
\newcommand\qfmean{\ensuremath{q_f}} %
\newcommand\qpmeanzxt{\ensuremath{\langle q_p\rangle_{zxt}}} %
\newcommand\qrefi{\ensuremath{q_{ref}}}    %
\newcommand\shields{\ensuremath{\theta}} %
\newcommand\shieldscrit{\ensuremath{\theta_c}} %
\newcommand\rhof{\ensuremath{\rho_f}} %
\newcommand\dratio{\ensuremath{\rho_p/\rho_f}} %
\newcommand\hfluid{\ensuremath{h_f}} %
\newcommand\hbed{\ensuremath{h_b}} %
\newcommand\hmean{\ensuremath{H_f}}
\newcommand\hbedmean{\ensuremath{H_b}}
\newcommand\hbedmeanphase{\ensuremath{\widetilde{H}_b}} %
\newcommand\yref{\ensuremath{y_0}} %
\newcommand\tbulk{\ensuremath{T_b}}
\newcommand\ufric{\ensuremath{u_\tau}}    %

\newcommand\ubulk{\ensuremath{u_b}}

\newcommand\phibed{\ensuremath{\Phi_b}}

\newcommand\phimeansmoothspanwise{\ensuremath{\langle \phi_p \rangle_z}}

\newcommand\ugrav{\ensuremath{U_{g}}}

\newcommand\meanlambda{\ensuremath{\lambda_{h}}}
\newcommand\criticallambda{\ensuremath{\lambda_{c}}}
\newcommand\minimumlambda{\ensuremath{\lambda_{th}}}

\newcommand\rmsamplitude{\ensuremath{\sigma_{h}}}
\newcommand\celerity{\ensuremath{u_{D}}}
\newcommand\tsep{\ensuremath{\delta t}}
\newcommand\rsep{\ensuremath{\delta x}}
\newcommand\corrx{\ensuremath{R_h}}

\newcommand\sheartotmean{\ensuremath{\langle \tau_{tot} \rangle}}

\newcommand\dpdxmean{\ensuremath{
  \Big\langle\frac{{\rm d}p}{{\rm d}x}\Big\rangle}} 
\newcommand\dpdxmeanline{\ensuremath{
   \langle{\rm d}p/{\rm d}x\rangle}}

%% file: abstract.tex
\begin{abstract}
The phenomenon of sediment pattern formation in a channel flow 
is numerically investigated 
by performing simulations which
resolve all the relevant length and time scales of the problem. 
The numerical approach employed and the flow configuration considered
is identical to our previous study 
(Kidanemariam and Uhlmann J.~Fluid~Mech., vol. 750, 2014, R2), 
the only difference being the length of the computational domain. The
latter was systematically varied in order to investigate its influence
on the initiation and evolution aspects.  By successively reducing the
streamwise length, the minimum box dimension which accommodates an
unstable sediment bed is revealed, thus determining the lower
threshold of the unstable modes. 
For the considered parameter point, the cutoff length 
for pattern formation
lies in the range $75$--$100$ times the particle diameter 
($3$--$4$ times the clear fluid height).
\revision{%
  Furthermore, the flow in a long streamwise box with a size of
  48 times the clear fluid height is simulated which  
  minimizes the domain
  size influence on the selection of the initial wavelength. It turns
  out that the longest box was able to accommodate approximately 11
  initial ripple units with wavelength in the range
  $100$--$110$ particle diameters.
  Note incidentally that this largest box contains well over one million 
  particles, which is by far the largest number of finite-size resolved
  particles which have ever been simulated in a turbulent flow over a
  long time interval.}{%
  We also simulate the flow in a very long streamwise box with a size of 48
  times the clear fluid height (featuring well over one million
  particles), accommodating approximately 11 initial ripple
  units with a wavelength in the range of 100--110 particle diameters.}
\revision{%
  The evolution of the amplitude of the patterns, 
  which was inferred from the auto-correlation of the 
  sediment bed height spatial fluctuation, 
  exhibits two regimes of growth: an initial exponential regime,
  with a growth rate independent of the chosen domain size 
  and a subsequent non-linear regime which is strongly constrained 
  by the domain length. }
{%
  The evolution of the amplitude of the patterns 
  exhibits two regimes of growth: an initial exponential regime,
  with a growth rate independent of the chosen domain size, 
  and a subsequent non-linear regime which is strongly constrained 
  by the domain length. 
}
In the small domain cases, after the initial exponential regime,
the ripples evolve steadily maintaining their
shape and migration velocity, at a mean wavelength equal to the 
length of the domain. The asymmetric ripple shape is
characterized by  a spectrum which exhibits 
a power-law decay over the first few dominant non-dispersive modes 
propagating at the mean dune migration velocity. 
The rate of particle transport and the mean interface shear stress
exhibited an increase with increasing ripple dimensions. Nevertheless,
the relationship between the two was observed to be 
\revision{%
  sufficiently }
{%
  approximately 
}
described by the 
\revision{%
  modified }\
{}
empirical power law formula for sediment 
transport by \citet{Wong2006}.
\end{abstract}

%% file: introduction.tex
\section{Introduction}
\label{sec:intro}
Sediment patterns, also commonly termed as dunes, ripples
or simply bedforms, can be abundantly observed in 
\revision{}{deserts, coastal areas,} 
natural streams or
man-made canals. These sediment features, besides being simply
fascinating, have important implications in many fields of science and
engineering. For instance, bedforms greatly influence the rate of
sediment transport as well as the stability of hydraulic structures in
a given river. Thus, fundamental understanding of the mechanisms which
are behind their formation as well as predicting their characteristics
is crucial. However, this task has been very challenging due to the
complex interaction between the sediment particles and the driving
turbulent flow.

\revisionold{%
  Sedimentary patterns could be broadly classified into aeolian and
  subaqueous. The former being driven by wind forces while the latter,
  which the present study is focused on, form under water or similar
  fluid. }
{%
  Sediment patterns in general can be broadly classified as 
  `aeolian' or `subaqueous', the former being driven by wind forces
  while the latter, which the present study is focused on, form under
  water or due to the action of a similar fluid. 
}
The sediment-to-fluid density ratio of subaqueous patterns 
is much smaller than that of the aeolian ones. 
As a result, although the basic formation process is the same, 
there are fundamental differences regarding the relevant
controlling mechanisms of their generation in both
categories
\citep[cf.][]{Bagnold1941,Sauermann2001,Andreotti2013,Duran2014}. 
\revisionold{%
  Furthermore,
  subaqueous patterns could be  categorized into those that form
  under the action of a steady flow, such as in rivers and
  canals, and those that are worked
  by an unsteady oscillatory flow, for instance by the action of a sea
  wave in coastal regions
  \citep[cf.][]{Sleath1976,Blondeaux1990,Vittori1990,Blondeaux2015}.
  The present work is further restricted to the study of
  patterns occurring in steady flows. 
  Strictly speaking, subaqueous bedforms, even those forming under
  steady flow conditions, are three-dimensional. 
  Nevertheless, under certain circumstances, bedforms tend to be statistically 
  invariant in the cross-stream direction
  \citep{Yalin1977}. 
}
{%
  
  Concerning subaqueous sediment patterns,
  one can further distinguish between 
  those that form 
  under the action of a steady flow, such as in rivers and
  canals, and those that are worked
  by an unsteady oscillatory flow, for instance by the action of a sea
  wave in coastal regions
  \citep[cf.][]{Sleath1976,Blondeaux1990,Vittori1990,Blondeaux2015}.
  Henceforth we will restrict our attention to 
  the study of patterns occurring in steady flows.  
  Although subaqueous bedforms are three-dimensional objects, 
  under certain circumstances they tend to be statistically 
  invariant in the cross-stream direction
  \citep{Yalin1977}. 
}
\revisionold{%
  Indeed, a large part of the previous theoretical modeling and 
  experimental measurement has focused on two-dimensional subaqueous bedforms
  \citep{Best2005}.  
  On the other hand, it is common to observe bedforms which are 
  highly three-dimensional affecting the mean flow in all spatial directions
  \citep{Allen1968,Best2005}.
  A particular type of three-dimensional subaqueous bedforms are the 
  `Barchan dunes' which are isolated features having a distinct 
  crescent like shape \citep{Groh2008,Franklin2009,Hersen2002}. 
  The present work focuses only on 
  transverse bedforms which are statistically
  invariant in the cross-stream direction. }
{%
  Indeed, a large part of the previous theoretical modeling and 
  experimental measurement has focused on two-dimensional subaqueous
  bedforms \citep{Best2005}, 
  and this is the approach we take in the present work. 
}
\revisionold{%
  Furthermore, it is common practice to classify bedforms 
  based upon their 
  dimensions and their dynamics \citep{Julien1998}. 
  Classically, a distinction is made among
  ripples, dunes and antidunes \citep{Engelund1982,Garcia2008}. 
  Ripples and dunes are similar,  exhibiting an asymmetric triangular-like 
  shape with a gentle slope on the upstream side of their crest and a steeper 
  slope on the downstream side. }
{%
  
  Bedforms can be characterized with respect to their  
  spatial dimensions and their dynamics \citep{Julien1998}. 
  Classically, a distinction is made between 
  ripples, dunes and antidunes \citep{Engelund1982,Garcia2008}. 
  Ripples and dunes exhibit some similarities, such as an asymmetric
  triangular-like shape with a gentle slope on the upstream side of
  their crest and a steeper slope on the downstream side. }
The principal distinction between ripples and 
dunes is the separation of length scales; the wavelength of
observed ripples is commonly believed to scale with the grain size,
and that of larger scale dunes is supposedly controlled by the flow
geometry, typically the flow depth \citep{Yalin1977}. 
Experiments show that at low flow rates small-scale undulations appear 
out of an initially flat sediment bed, 
\revision{}{at first} 
having no clear-cut shape 
\citep{Coleman1994}. 
When increasing the flow rate, these ripples give way to the much larger
dunes, which are characterized by flow separation and recirculation
in their downstream faces, and which propagate downstream. 
At even higher flow rates antidunes appear, 
which tend to propagate upstream. In the case of a flow with a free surface, 
both dunes and antidunes dynamically interact with  surface waves
\citep[see e.g.][]{Best2005}. 
Nevertheless, bedforms are 
equally observed in closed conduits which possess no free surface, 
and thus the existence of a free surface is not necessary for their formation 
\citep{Yalin1977,Bridge1988,Langlois2007a,Ouriemi2009b,CardonaFlorez2016}. 
The problem of bedform and free surface wave interaction is not 
within the scope of the  present study.

In most previous theoretical work on pattern formation,
the background turbulent flow is typically represented
by a Reynolds averaged Navier-Stokes equations (RANS) 
model, while the  sediment bed evolution 
is described by the sediment continuity equation, i.e.\ the Exner equation.
The hydrodynamic and morphodynamic problems are then coupled by an algebraic
expression 
for the particle flux as a function of the local bed shear stress.  
Hydro-morphodynamic linear (or weakly nonlinear) 
stability analysis is then performed in order to determine the stability of the
sediment bed, the controlling parameter of the instability as well as the 
initial unstable pattern wavelength
\citep[e.g.\ see reviews by][]{Seminara2010,Charru2013a}. 
However, there is no clear consensus 
among the different predictions from these approaches, and, when compared to 
experimental observations, the outcome of these models is often unsatisfactory.
For instance, the prediction for the initial pattern wavelength can
be off by an order of magnitude
\citep{Raudkivi1997,Langlois2007a,Coleman2009,Ouriemi2009b}. 
This inadequacy can be linked to, among others, 
the insufficient predictive capability of the adopted algebraic expressions 
for the particle flux. 

The subsequent bedform evolution is an even more complicated issue
which includes aspects like coarsening (bedform wavelength and
amplitude growth), asymmetry, coalescence (fusion of bedforms), 3D patterning
(loss of bedform translational invariance). Even in simplified and
controlled experiments where bedforms are essentially two-dimensional
\citep[see e.g.][]{Betat2002a}, the evolution process is a
result of nonlinear interaction mechanisms which are far
from our understanding.  Certainly, linear stability theories are not
adequate in describing the bedform transient processes such as the
evolution of the amplitude or the evolution of bedform morphology
towards the asymmetric shape (which is even observed in the absence of
flow separation).  
\revisionold{%
  Weakly nonlinear stability theories have been
  proposed in some publications \citep[see e.g.][]{Colombini2008a}, but
  still are unable to predict or describe bedforms such as the ``vortex
  dunes'' observed in the experiments of \citet{Ouriemi2009b}. The
  vortex dunes are characterized by flow separation and recirculation at
  their fronts and their initial evolution dynamics differ from those of
  the small scale bedforms without flow separation
  \citep{Ouriemi2009b}. }
{%
  Weakly nonlinear stability theories have been
  proposed \citep[see e.g.][]{Colombini2008a}, but
  this approach is still unable to reliably predict the observed
  bedforms in reference experiments \citep[e.g.][]{Ouriemi2009b}. }

There exists a very limited amount of experimental
studies available reporting on the initial formation and development
stages of patterns
\citep[see e.g.][]
{Coleman1994,Betat2002a,Coleman2003,Langlois2007a,Ouriemi2009b,
CardonaFlorez2016}. 
Even these studies, limited by measurement 
difficulties, often fall short of  providing details with respect to 
the very first instants of the bed instability, such as the 
dispersion relation of the unstable modes. Such 
quantities are crucial  when it comes to assessing the validity 
of the various proposed theoretical models.
On the other hand, although there are a number of numerical studies
available on the problem of subaqueous sedimentary patterns, 
most of these studies are based on the continuum description of 
the sediment bed, which again rely on the use of
\mbox{(semi-)empirical} relationships to couple the hydro-morphodynamic problem
\citep[see e.g.][]{Chou2010,Khosronejad2014}. 
\revision{}{In the granular flow community,
a number of studies have been performed with the aid of a Lagrangian 
description of the sediment bed (based on discrete-element models), but 
without coupling to a resolved turbulent background flow
\citep{Duran2012,Schmeeckle2014,Maurin2015}}. 

There has been a promising gain of
momentum in recent years on the number of direct numerical simulation
(DNS) studies 
on the problem of sediment transport in which the 
detail of the flow, even at the boundaries of each individual 
particles, is faithfully resolved 
\citep[see e.g.][]{Kidanemariam2014b,Kidanemariam2014,
Vowinckel2014,Derksen2015}. 
We have recently performed novel DNS of the flow over an 
erodible bed of spherical 
sediment particles in a statistically unidirectional channel 
flow configuration
in both the laminar and turbulent regimes
\citep[][]{Kidanemariam2014}.  
These simulations are, to the best of our knowledge, the first to
successfully simulate the evolution of a bed of mobile sediment
particles (leading to pattern formation) by means of 
DNS \citep{Colombini2014}.  The results
from these simulations have shown that the chosen streamwise length
of the computational domain ($12\hmean$, where $\hmean$ is the channel fluid
height) was able to accommodate a few integer multiples of the initial pattern
wavelength, which allowed us to address some of the outstanding questions on
sedimentary patterns such as the initial dominant wavelength and
pattern amplitude growth \citep[][]{Kidanemariam2014}. Ideally, in
order to accurately capture the natural selection mechanism of the
unstable wavelength and its subsequent evolution, it
is necessary to consider a computational domain size which is much
superior than the anticipated pattern wavelength.  However, 
the domain sizes considered by
\citet{Kidanemariam2014}
might be marginal when compared to the
observed dominant wavelengths. It was noted that the discreteness of
the numerical harmonics  could influence the initial
wavelength as well as the time evolution processes.
In light of this aspect, assessing the influence of the domain size,
on the selection and evolution process is crucial.

Moreover, various theoretical
and experimental studies have shown that there is a lower threshold of  
the unstable wavelengths 
\citep[see for instance][]{Charru2006c,Franklin2011}.
Determination of this value is important, since the
lower bound of unstable modes is believed to be linked to some relevant
length scale which controls erodible bed instability 
\citep{Hersen2002,Claudin2006,Andreotti2011}. 

In the present work, we have carried out a series of simulations
of the same flow configuration as in our  
previous study \citep{Kidanemariam2014}. We have kept 
all the physical and numerical parameters of the simulations  
identical, except for the streamwise length of the computational box.
The latter was varied between $3\hmean$ and $48\hmean$,
in order to determine whether at 
the cosidered parameter point, 
there exists a  cutoff length for pattern formation. 
To this end, we have successively reduced the domain size
until it is smaller than an unknown threshold and cannot accomodate
an unstable bed. Moreover, in order 
to assess the influence of the domain size on the selection of the 
initial ripple wavelength and its subsequent evolution,
we have chosen a computational domain size which is four times 
larger than in our previous study. It should be noted that  
a very large number of spherical particles (approximately 
1.1 million in total) are considered to represent the mobile bed.
This simulation is the first of its kind to break the $\mathcal{O}(10^6)$ fully
resolved particle milestone.
Furthermore, based on the analysis of the DNS data, we address the
relationship between the evolving patterns, their migration velocity and
the particle flow rate \revision{}{at the present parameter point}.  

%% file: numerics.tex
\section{Numerical method}
\label{sec:numerics}

In the present work we have used the same numerical procedure as
\citet{Kidanemariam2014,Kidanemariam2014b}.
The numerical treatment of the fluid-solid system is based upon the
immersed boundary technique of \citet{Uhlmann2005a}, 
wherein the incompressible Navier-Stokes equations are solved with a
second-order finite-difference method throughout the
entire computational domain $\Omega$, adding a localized force term
which serves to impose the no-slip condition at the fluid-solid
interface.  
The particle motion is obtained via integration of the Newton-Euler
equations for rigid body motion, driven by the hydrodynamic force 
(and torque) as well as gravity and the force (torque)
resulting from solid-solid contact. 
The collision process between the immersed particles is described
through a discrete element model (DEM) based on the soft-sphere
approach. 
A pair of particles is defined as `being in contact' 
when the smallest distance between their surfaces, $\Delta$, becomes
smaller than a force range $\Delta_c$. 
The resulting contact force is then the sum of an elastic normal
component, a normal damping component and a tangential frictional
component. 
The elastic part of the normal force component 
is a linear function of the penetration length  
$\delta_c \equiv \Delta_c-\Delta$, with a stiffness constant $k_n$. 
The normal damping force is a linear function of the normal component 
of the relative velocity between the particles at the 
contact point with a constant coefficient $c_n$. 
The tangential frictional force (the magnitude of which is limited 
by the Coulomb friction limit with a friction coefficient $\mu_c$) is
a linear function of the tangential 
relative velocity at the contact point, again formulated with a
constant coefficient denoted as $c_t$. 
A detailed description of the collision model
and extensive validation can be found in \citet{Kidanemariam2014b}.  

The four parameters which describe the collision process in the
framework of this model ($k_n$, $c_n$, $c_t$, $\mu_c$) as well as the
force range $\Delta_c$ need to be prescribed for each simulation.  
Note that the normal stiffness coefficient $k_n$ and the normal
damping coefficient $c_n$ can be related by introducing the dry
restitution coefficient $\varepsilon_d$, defined as the absolute 
value of the ratio between the normal components of the relative
velocity post-collision and pre-collision. 

In the present simulations, $\Delta_c$ is set equal to 
one grid spacing $\Delta x$.  
The stiffness parameter $k_n$ has a value equivalent to approximately 
$17000$ times the submerged weight of the particles, divided by
the particle diameter. 
The chosen value ensures that the maximum overlap $\delta_c$ over
all contacting particle pairs is within a few percent of
$\Delta_{c}$.  
The dry coefficient of restitution is set to $\varepsilon_d=0.3$ 
which together with $k_n$ fixes the value for $c_n$. 
Finally, the tangential damping coefficient $c_t$ was set equal to
$c_n$, and a value of $\mu_{c}=0.4$ was imposed for the Coulomb
friction coefficient.   
This set of parameter values for the contact model is the same as used
by \citet{Kidanemariam2014b}. 

Since the characteristic collision time is typically orders of
magnitude smaller than the time step of the flow solver, the
numerical integration of the equations for the particle motion is
carried out adopting a sub-stepping technique,
freezing the hydrodynamic forces acting upon the particles between
 successive flow field updates \citep{Kidanemariam2015}.  

%% file: setup.tex
\section{Flow configuration and parameter values}
\label{sec:setup}
\begin{figure}%
\begin{minipage}{\linewidth}
  \centering
  \includegraphics[width=0.6\linewidth,clip=true,viewport=100 465 350 645]
           {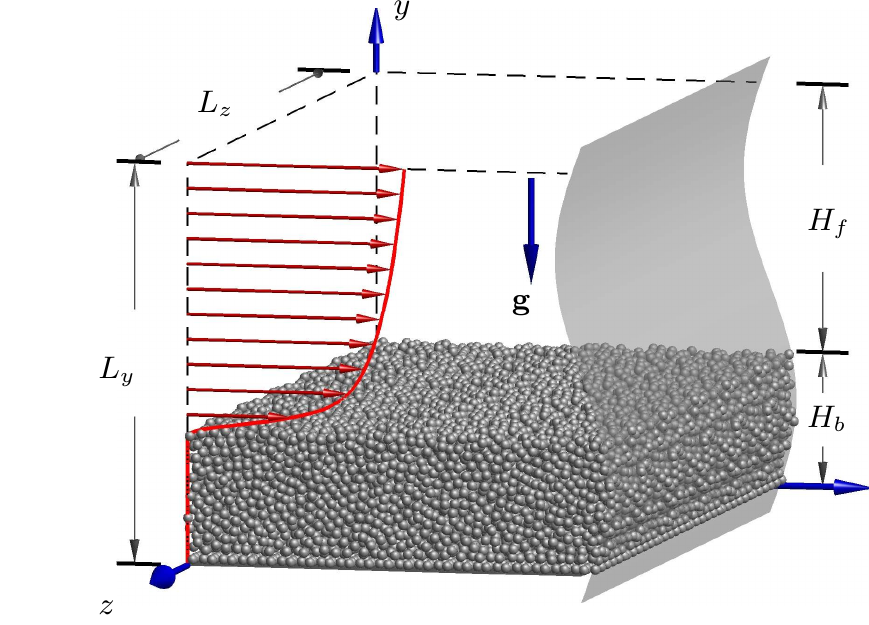}
\end{minipage}
\caption{Schematic diagram showing the configuration of
            the simulations.
            The computational domain %
            is periodic
            along the $x$- and $z$-directions. The no-slip and 
            free-slip boundary conditions are imposed at the   
            bottom ($y=0$) and top ($y=\Ly$) respectively.
            }
   \label{fig:schematic-diagram}
\end{figure}
We have performed a total of ten independent
simulations of the
development of bedforms over a subaqueous sediment in an open channel
flow configuration.
As shown in figure \ref{fig:schematic-diagram} a Cartesian coordinate
system is adopted such that $x$, $y$, and $z$ are the streamwise,
wall-normal and spanwise directions, respectively. Mean flow and
gravity are directed in the positive $x$ and the negative $y$
directions respectively. The computational domain is periodic in the
streamwise and spanwise directions.  A free-slip condition is
imposed at the top boundary while a no-slip condition is imposed at
the bottom wall.    The simulations are labeled \caseA, \caseB,
\caseC, \caseD, \caseE\ and \caseF, indicating the approximate
streamwise box length in terms of the mean fluid height \hmean.
The mean fluid height \hmean\ and the corresponding mean sediment
bed thickness \hbedmean\ are computed by performing streamwise and
time averaging of the spanwise-averaged instantaneous fluid height \hfluid\
and sediment bed height \hbed\
(note that the definition of \hfluid\ and \hbed\ will be made more precise
in section \ref{subsec:fluid-bed-interface}).  In order
to perform ensemble averaging, case~\caseB\ and
\caseE\ are performed three times each adopting different initial
conditions.
Each simulation is performed independently following the simulation
start-up procedure as detailed in \citet{Kidanemariam2014}.

In all cases, the channel
is driven by a horizontal mean pressure gradient which is adjusted at 
each time step in order to impose a constant flow rate
\qfmean. This results in a shearing flow of fluid height \hmean\ 
over a mobile bed of height \hbedmean\
(cf.\ sketch in figure \ref{fig:schematic-diagram}).
As is shown in table~\ref{tab:physical-parameters}, 
the bulk Reynolds number of the flow, which is defined based on the 
mean height
of the fluid as 
\begin{equation}\label{eq:bulk-reynolds-number}
   \Reb = \frac{\ubulk\hmean}{\nu},
\end{equation}
where $\ubulk\equiv\flowrate/\hmean$ is the bulk velocity 
and $\nu$ is the kinematic viscosity, is set at a value such that the 
flow is fully turbulent. 
The friction Reynolds
number \Ret, which is similarly defined 
based on the friction velocity \ufric, is {\it a posteriori} determined
by  evaluating the total shear stress 
at the wall-normal location of the mean fluid-bed interface
$y=\hbedmean$ (cf.\ section 
\ref{subsec:bottom-friction} for details of the
determination of \ufric).

\begin{table}
   \begin{center}
  \begin{tabular}{llrrrrrrr}
     Case &  \Reb\ & \Ret\  & \dratio\ & \Ga & $D^+$ & 
     $\hmean/D$ & $\hbedmean/D$ & $\shields$ \\  [6pt]
\input{patt_physical_paramters_table_modif.tex}
    \end{tabular}
     \caption{\label{tab:physical-parameters}
           Physical parameters of the simulations.
           In all cases, the values of the bulk Reynolds number 
           \Reb, the particle-to-fluid density ratio 
           \dratio\ and the Galileo number \Ga\ are imposed. 
           Derived physical parameters include
           the friction Reynolds number \Ret,
           the Shields number \shields\ as well as the length scale ratios
            $\hmean/D$,  $\hbedmean/D$ and $D^+$. 
           Note that three independent simulations are performed for 
           cases \caseB\ and \caseE\ each for the purpose of ensemble
           averaging. 
           The values of the relevant parameters are determined 
           \revision{a given time interval at the end of the corresponding 
           simulation period}
           {over a time interval at the end of the
            respective simulations}        
           (cf.\ table~\ref{tab:numerical-parameters}).
          }
  \end{center}
\end{table}
\begin{table}
  \begin{center}
  \begin{tabular}{lrrrrrr}
     Case & $[\Lx \times \Ly \times \Lz]/D$ & $D/\Delta x$ & 
     $\Delta x^+$ & \Npt & $T_{obs}/\tbulk$  
      & $T_{obs}^{s}/\tbulk$ \\  [6pt]
   \caseA & $76.8\times 38.4\times 76.8$  & 10 & 0.96 & 65359  & 
     \revision{}{638} &  \revision{}{483}\\
   \caseB\textsuperscript{1,2,3} 
   & $102.4\times 38.4\times 76.8$ & 10 & 1.06 & 86645  & 
        401/400/\revision{}{853} & 
        88/84/\revision{}{542}\\
   \caseC & $153.6\times 38.4\times 76.8$ & 10 & 1.19 & 127070 & \revision{}{918} & 
        \revision{}{513}\\
   \caseD & $179.2\times 38.4\times 76.8$ & 10 & 1.22 & 150521 & \revision{}{977} & 
        \revision{}{566}\\
   \caseE\textsuperscript{1,2,3} 
   & $307.2\times 38.4\times 76.8$ & 10 & 1.20 & 263412 & 911/807/890 & 
        283/177/262\\
   \caseF & $1228.8\times 38.4\times 76.8$& 10 & \revision{}{1.17} & 1053648& 
       \revision{}{462} & 
        \revision{}{28}\\

    \end{tabular}
     \caption{\label{tab:numerical-parameters}
           Numerical parameters of the simulations.
           $L_i$ is the domain
           length adopted in the $i$-th direction. In all the cases,
           the spherical
           particles have a diameter $D=10/256$ and a uniform grid
           spacing $\Delta x = \Delta y = \Delta z =1/256$ 
           is adopted yielding the above
           listed resolutions. \Npt\ is the number of spherical
           particles adopted. $T_{obs}$ is the total simulation
           time interval staring from the instant at which the mobile particles 
           are released, while $T_{obs}^{s}$ corresponds
           to the interval
          \revision{}{(at the end of the respective simulations)}
          during which statistical averaging 
           is performed to   
           determine the relevant parameters listed in 
           table~\ref{tab:physical-parameters} 
           \revision{}{(\Ret, $D^+$, $\hbedmean/D$, and $\shields$)}
           and to compute 
           steady-state ripple quantities which are discussed in 
           sections \ref{subsec:dune-morphology}, 
            \ref{subsec:propagation-speed},
            \ref{subsec:bottom-friction} and \ref{subsec:particle-flow-rate}. 
           $\tbulk\equiv \hmean/\ubulk$ is the bulk time unit.
           }
   \end{center}
\end{table}

In order to fully describe the flow, at least 
three parameters in addition to \Reb\ need to be imposed: the 
particle-to-fluid density ratio \dratio, a length scale ratio
$\hmean/D$, and a ratio between the gravity and viscous forces
which is described by the Galileo number viz. 
\begin{equation}\label{eq:galileo-number}
\Ga = \frac{((\dratio-1)|\mathbf{g}|D^3)^{\frac{1}{2}}}{\nu} 
    = \frac{U_gD}{\nu}\;,
\end{equation}
where $\mathbf{g}$ is the gravitational acceleration.
$Ga$ can be considered as a Reynolds number defined based on the 
gravitational
\revision{settling velocity}{velocity scale} 
$U_g=\sqrt{(\dratio-1)|\mathbf{g}|D}$
and the particle diameter.
\revision{}{Note that $U_g$ is not the actual settling velocity 
  \citep{Jenny2004}. 
}
The density ratio is fixed at a value  of 
$\dratio=2.5$ which corresponds to the density ratio between glass 
beads and water.  
The thickness 
of the fluid height and that of the erodible bed are chosen to be
sufficiently large  for the formation of the anticipated bed patterns. 
This results in a very large number
of spherical particles due to the size the computational box
(cf.\ table~\ref{tab:numerical-parameters}). 
Finally, the value of the Galileo number is chosen by adjusting the 
value of acceleration due to gravity $|\mathbf{g}|$, such that
the value of the Shields number 
\begin{equation}\label{eq:shields-number}
  \shields =
  \frac{\ufric^2}{\left(\dratio-1\right)|\mathbf{g}|D}
  = \frac{\ufric^2}{U_g^2}
  \revision{}{ = \left(\frac{D^+}{\Ga}\right)^2}
    \,,
\end{equation}
is above the critical value  \shieldscrit\ for incipient sediment 
motion.
In the turbulent flow regime, \shieldscrit\ is believed to 
\revision{%
  be approximately $0.05$ for 
  $D^+\equiv D\ufric/\nu=\mathcal{O}(1)$ citep{Wong2006,Franklin2011} }
{%
  approximately lie in the range $0.03\ldots0.05$ with a mild
  dependence upon the Galileo number
  \citep{Wong2006,Franklin2011,soulsby:97}.  
}

\revisionold{%
  We remark that, as has been shown by \citet{Kidanemariam2014}, 
  the adopted parameter point falls in the 
  `vortex dune' regime classification of the pipe flow experiment
  by \citet{Ouriemi2009b}. 
  }
  {%
    We remark that the present parameter point is located in the 
    `vortex dune' regime in the classification of \citet{Ouriemi2009b}. 
    These authors coined the term `vortex dunes' for patterns which are
    characterized by flow separation and recirculation at their
    downstream face, in order to distinguish them from
    smaller-amplitude patterns without flow separation which they
    observed in their pipe-flow experiments (the latter were termed  
    'small dunes' therein). Note that this terminology should not be
    confused with the distinction betwen `ripples' and `dunes'
    commonly made in the river flow community;
    in particular, it does not imply that the length scales of the
    present patterns exhibit a scaling with the water depth. 
  }

\revision{}
{%
  Incidentally, let us mention that the chosen parameters of our
  simulation can be transformed into a physically realizable
  laboratory experiment. 
  For instance, 
  choosing as working fluid a mixture of 94\% pure water and 6\%
  UCON oil 75H-90000 (leading to kineamatic viscosity of the mixture
  of $4.3\times10^{-6} m^2/s$ and 
  a mixture density of $1002 Kg/m^3$), and selecting $D=1mm$ glass
  beads implies the following dimensions of the experimental apparatus
  (under terrestrial conditions): 
  channel height $3.9cm$, fluid height $2.5cm$,
  wavelength of obtained ripples in the range of $10cm\ldots18cm$. 
  The bulk velocity in this hypothetical set-up would measure
  $0.52m/s$. 
}

%% file: patt_physical_paramters_table_modif.tex
\caseA  & 3011 & \revision{}{244.8} & 2.5 &28.37 & \revision{}{9.62}  & \revision{}{25.44} & \revision{}{12.96} & 0.12\\
\caseBa & 3011 &             273.6  & 2.5 &28.37 &             10.89  &             25.12  &             13.28  & 0.15\\
\caseBb & 3011 &             265.4  & 2.5 &28.37 &             10.55  &             25.15  &             13.25  & 0.14\\
\caseBc & 3011 & \revision{}{263.0} & 2.5 &28.37 & \revision{}{10.45} & \revision{}{25.17} & \revision{}{13.23} & 0.14\\
\caseC  & 3011 & \revision{}{303.0} & 2.5 &28.37 & \revision{}{11.90} & \revision{}{25.47} & \revision{}{12.93} & 0.18\\
\caseD  & 3011 & \revision{}{309.1} & 2.5 &28.37 & \revision{}{12.22} & \revision{}{25.30} & \revision{}{13.10} & 0.19\\
\caseEa & 3011 &             301.3  & 2.5 &28.37 &             12.01  &             25.08  &             13.32  & 0.18\\
\caseEb & 3011 &             301.6  & 2.5 &28.37 &             12.04  &             25.05  &             13.35  & 0.18\\
\caseEc & 3011 &             298.6  & 2.5 &28.37 &             11.91  &             25.07  &             13.33  & 0.18\\
\caseF  & 3011 & \revision{}{293.1} & 2.5 &28.37 & \revision{}{11.69} & \revision{}{25.07} & \revision{}{13.33} & \revision{}{0.17}\\

%% file: dune_dimensions.tex
\section{Extraction of bedform dimensions}
\label{sec:dune-dimensions}
\subsection{Definition of the fluid-bed interface}
\label{subsec:fluid-bed-interface}
The details of the extraction of the fluid-bed interface, which
has been given in our previous work \citep{Kidanemariam2014,Kidanemariam2015}, is
briefly repeated here for completeness.
The location of the interface between the fluid and the sediment bed
has been determined based on the threshold value of the solid
volume fraction in the following way. 
First, a solid phase indicator function $\phi_p(\mathbf{x},t)$ is
defined which 
has a value of one if $\mathbf{x}$ is located inside any
particle and zero elsewhere. 
Spanwise averaging then yields 
$\phimeansmoothspanwise(x,y,t)$ which is a direct measure of the 
instantaneous, two-dimensional solid volume fraction. 
The spanwise-averaged fluid-bed interface location $\hbed(x,t)$ is
finally extracted by means of a threshold value, chosen as
$\phimeansmoothspanwise^{thresh} = 0.1$ 
\citep{Kidanemariam2014b},  
viz.
\begin{equation}\label{eq:bed-height} 
   \hbed(x,t)  = y \;\; \vert \; \phimeansmoothspanwise(x,y,t)
                                =\phimeansmoothspanwise^{thresh}\;.
\end{equation}
The corresponding spanwise-averaged fluid height is then simply given
by 
\begin{equation}\label{eq:fluid-height} 
   \hfluid(x,t)=\Ly - \hbed(x,t).  
\end{equation}

In the present study, we infer 
the evolution of the dimensions of the patterns as well as their 
two-dimensional shape and propagation velocity by scrutinizing the 
spatial and temporal variation  of the sediment bed height \hbed\
or its fluctuation 
with respect to the instantaneous average bed height,  
\begin{equation}\label{eq:bed-height-fluctuation}
  \hbed^\prime(x,t) =  \hbed(x,t) - \langle \hbed \rangle_x(t)\;.
\end{equation}
\subsection{Definition of pattern amplitude and wavelength}
\label{subsec:amplitude-and-wavelength}
The most relevant parameters, among others, which describe the geometrical
features of statistically two-dimensional bedforms are their streamwise and
wall-normal dimensions, namely the mean wavelength and mean amplitude of the
patterns.
As reviewed by
\citet{Coleman2011}, various definitions have been used in order to
quantify these dimensions. For instance, the average wall-normal distance
between local maxima and adjacent local minima of \hbed\ is used as a measure of
the amplitude of the patterns in some studies \citep[][]{Ouriemi2009b}. Other
studies infer the amplitude of the patterns statistically from 
the r.m.s.\ fluctuation of
the bed \citep[][]{Langlois2007a}.
Similarly, the wavelength of patterns is quantified either geometrically, 
for instance based on the mean spacing between alternating
troughs or ridges of the sediment bed height, or statistically, for instance
based on the autocorrelation of the bed height fluctuation \citep{Coleman2011}.
In the present study, we have adopted a statistical definition.

The wall-normal dimension of the patterns is characterized
by the r.m.s. fluctuation of the sediment bed \rmsamplitude, 
defined as
\begin{equation}\label{eq:rms-bed-height}
 \rmsamplitude^2(t) 
   = \langle \hbed^\prime(x,t)\cdot\hbed^\prime(x,t)
     \rangle_x\;.%
\end{equation}
In order to determine the average wavelength of the patterns, first 
we define the instantaneous two-point correlation coefficient of 
the bed height fluctuation  
as a function of streamwise separation \rsep\ as 
\begin{equation}\label{eq:two-point-auto-correlation}
   \corrx(\rsep,t) = 
     \frac{
    \langle \hbed^\prime(x,t)\cdot\hbed^\prime(x+\rsep,t)\rangle_x
     }{\rmsamplitude^2(t)}\;.
\end{equation}
At a given time instant $t=t_1$, the correlation coefficient
$\corrx(\rsep,t_1)$ of a sediment bed featuring 
\revision{ripples or dunes}{bedforms},
exhibits a distinct damped-oscillation curve featuring alternating positive 
and negative values as a function of \rsep. 
Then, we  define an average bedform wavelength \meanlambda\ as twice the 
streamwise separation $\rsep=x_{min}$ at 
which the global minimum of \corrx\ occurs viz.
\begin{equation}\label{eq:mean-wavelength}
     \meanlambda(t) = 
              2x_{min} \;|\; \forall\, \rsep\in [0,\,\Lx/2] :
              \corrx(\rsep,t)\ge \corrx(x_{min},t)\,.
\end{equation}

%% file: result.tex
\section{Results}
\label{sec:results}
During the preparation of the present manuscript we have discovered a
programming error in the solid contact part of the simulation
code. This bug was active during the simulations of
\citet{Kidanemariam2014}. After correcting the error, we have rerun
those simulations and compared the results. This comparison is discussed in
appendix~\ref{sec:appendix-old-vs-new-data}. The conclusion from this assessment
is that the results did not change qualitatively.
Concerning the quantitative comparison, it can be stated that 
the difference between the result with
and without bug is within the scatter due to the three independent
realizations simulated here. 
\subsection{The minimum and the most amplified 
            unstable pattern wavelengths}
\label{subsec:unstable-pattern-wavelength}
\begin{figure}
   \centering
   \includegraphics[width=\textwidth,clip=true,viewport=70 150 520 750]
           {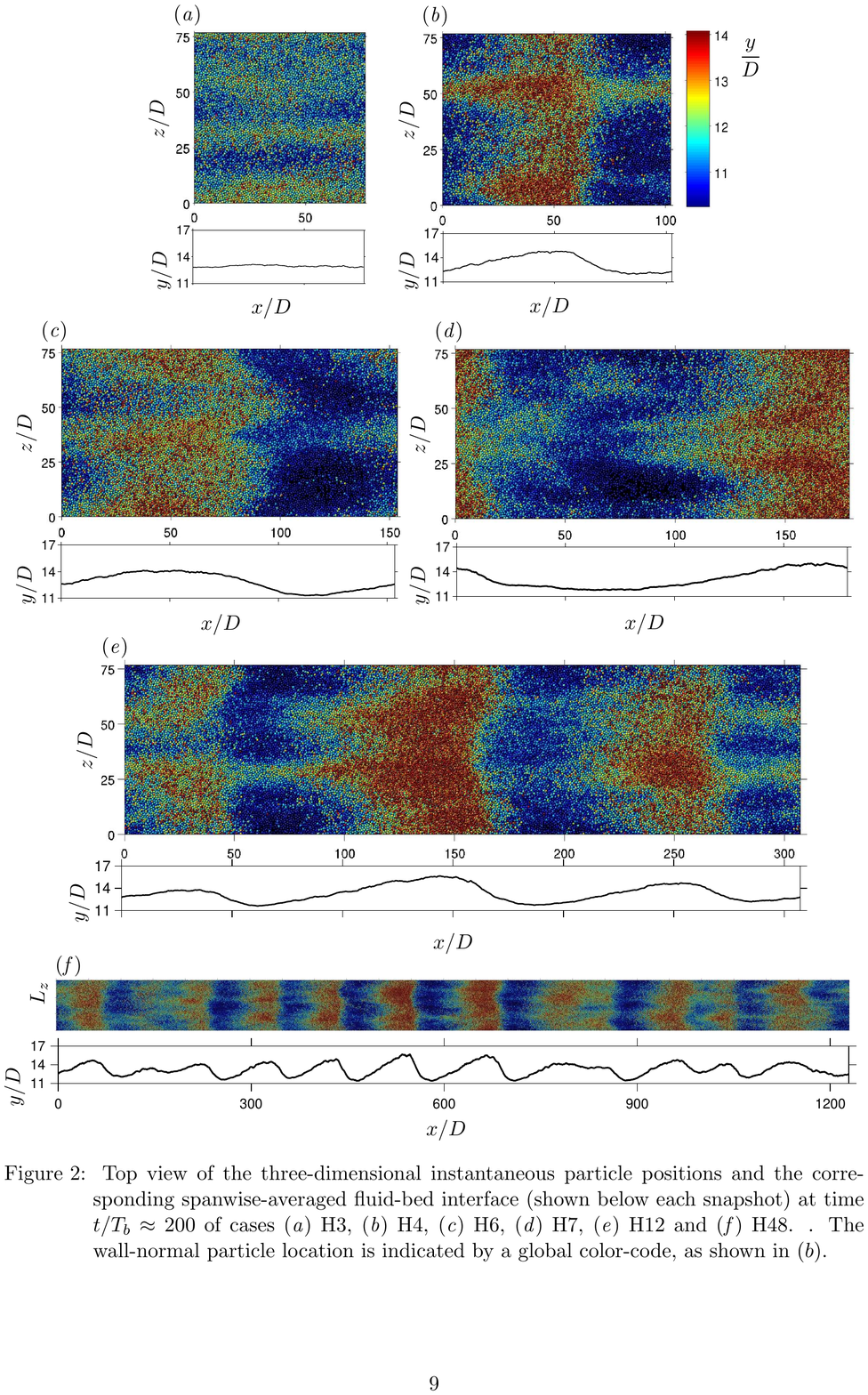}
\caption{
     Top view of the three-dimensional instantaneous 
     particle positions 
     \revision{}{and the corresponding spanwise-averaged
       fluid-bed interface (shown below each snapshot)}
     at time $t/\tbulk\approx200$ of cases
     (\textit{a})~\caseA,
     (\textit{b})~\caseB,
     (\textit{c})~\caseC,
     (\textit{d})~\caseD,
     (\textit{e})~\caseE\ and
     (\textit{f})~\caseF. 
     \revision{showing only the fourth quarter of the 
       domain
       ($921.6 < x/D < 1228.8$) in order to allow
       a direct visual comparison to the other cases.
       (\textit{g}) case~\caseF\ (covering the full domain)}{}
     \revision{%
       The wall-normal  dimension is visualized by 
       coloring each particle according to its wall-normal location
       (increasing from blue to red).}
     {%
       The wall-normal particle location is indicated by a global
       color-code, as shown in $(b)$.  
     }
     \protect\label{fig:instantaneous-particle-snapshots}
   }   
 \end{figure}

Our strategy to find the minimal box length \Lx\ 
which will accommodate the lower 
bound of the unstable wavelengths \minimumlambda\ (at a given parameter
point) is to perform a series of
numerical experiments in which the streamwise domain length 
is successively reduced. 
The concept is similar to the minimal flow unit of 
\citet{Jimenez1991}. Below a threshold value of $\Lx=\minimumlambda$,
pattern evolution will be hindered and a perturbed bed should in 
principle be stable even though it is in a regime where instability is 
expected.
Note that determining the lower threshold of \minimumlambda\ 
for bed instability 
does not necessarily mean determining the most amplified wavelength 
\criticallambda. The latter, if it uniquely exists,
 exhibits the maximum growth rate 
of the bed instability.
Figure~\ref{fig:instantaneous-particle-snapshots} shows
instantaneous snapshots of the particle positions in the
different cases after approximately 200 bulk time units have 
elapsed. 
It can be seen that the sediment bed of case~\caseA\ does not
feature  dune-like patterns. However, one can observe, 
that it exhibits distinct streamwise
aligned alternating ridges and troughs, although the sediment bed is
essentially flat when 
the spanwise average of the sediment bed profile is observed.
Such an organization of particles is linked to the
streamwise-aligned near-wall turbulent structures (streaks) which are
responsible for the non-homogeneous spatial distribution of particles.
The highly regular ridge-trough pattern in this case is a result of the
very small box length which hinders spatial de-correlation of the
turbulent structures in the streamwise direction.
On the other hand, the sediment bed of the remaining cases 
with $L_x \geq 4\hmean$ is seen to 
be unstable allowing for  the formation of spanwise-oriented patterns 
(the streamwise-aligned patterns are also visible superimposed to the
\revision{}{ripples}\footnote{In the remainder
of the present text we will refer to the observed bedforms as ripples.}). 
This indicates that the value of \minimumlambda\ lies
somewhere between the domain lengths of cases \caseA\ and \caseB. Another
observation is that the initial wavelength of the emerging patterns 
seems to be different depending on the chosen domain length.
In particular, the case with the largest domain size, 
which has a  
streamwise box length $12$ times the smallest domain length of 
the ripple-featuring cases,
is observed to accommodate more than ten initial ripple units. The
dimension of the ripples in the large domain case seems to have 
evolved less constrained by the 
box size when compared to those in the small domain cases.
Let us provide a more quantitative analysis of the above 
observation by evaluating the r.m.s. sediment bed height fluctuation
\rmsamplitude\ of each case and using it as a criterion to determine whether a 
sediment bed in a given simulation is stable or not. 
That is, if \rmsamplitude\ 
remains bounded within
a certain threshold value,
then the bed is said to be stable. On the other hand, 
for an unstable bed, \rmsamplitude\ will initially exhibit a 
continuous increase as a function of time.
Note that, even a stable flat (mobile) sediment bed features a small but finite
value of  \rmsamplitude\ as a result of the random uncorrelated 
bed undulations which stem from the discreteness of the bed at the grain 
scale. For instance, \citet{Coleman2009} report $\rmsamplitude/D\approx0.17$
for their macroscopically ``flat'' bed.  

\begin{figure}
   \centering
        \begin{minipage}{3ex}
          \rotatebox{90}
          {\small \hspace{2ex}$\rmsamplitude/D$}
        \end{minipage}
        \begin{minipage}{.45\linewidth}
          \centerline{(\textit{a})}
          \includegraphics[width=\linewidth]
          {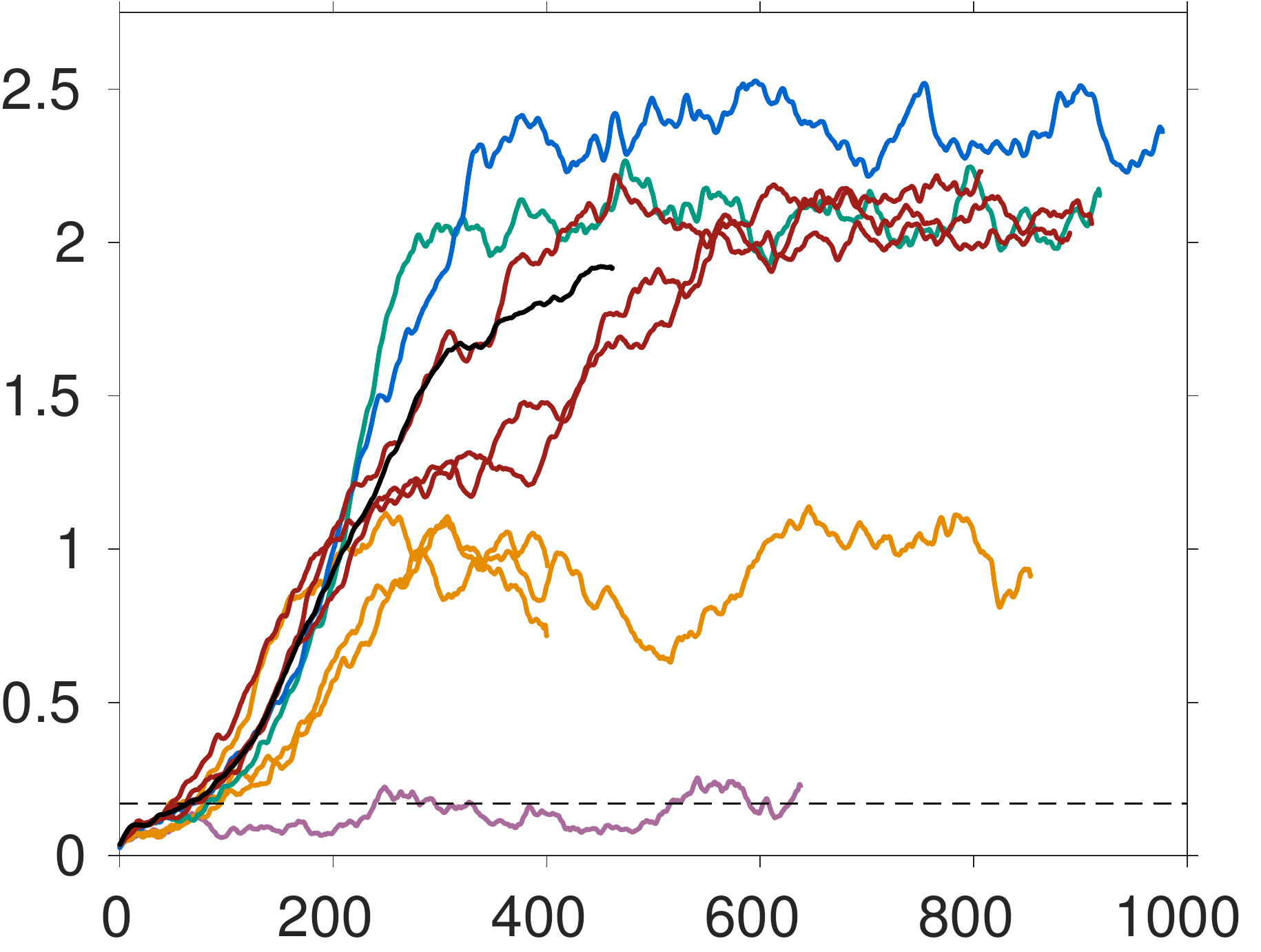}
          \centerline{\small  $t/\tbulk$}
        \end{minipage}%
         \hfill
         \begin{minipage}{3ex}
           \rotatebox{90}
           {\small \hspace{2ex}$\rmsamplitude/D$}
         \end{minipage}
         \begin{minipage}{.45\linewidth}
           \centerline{(\textit{b})}
           \includegraphics[width=\linewidth]
           {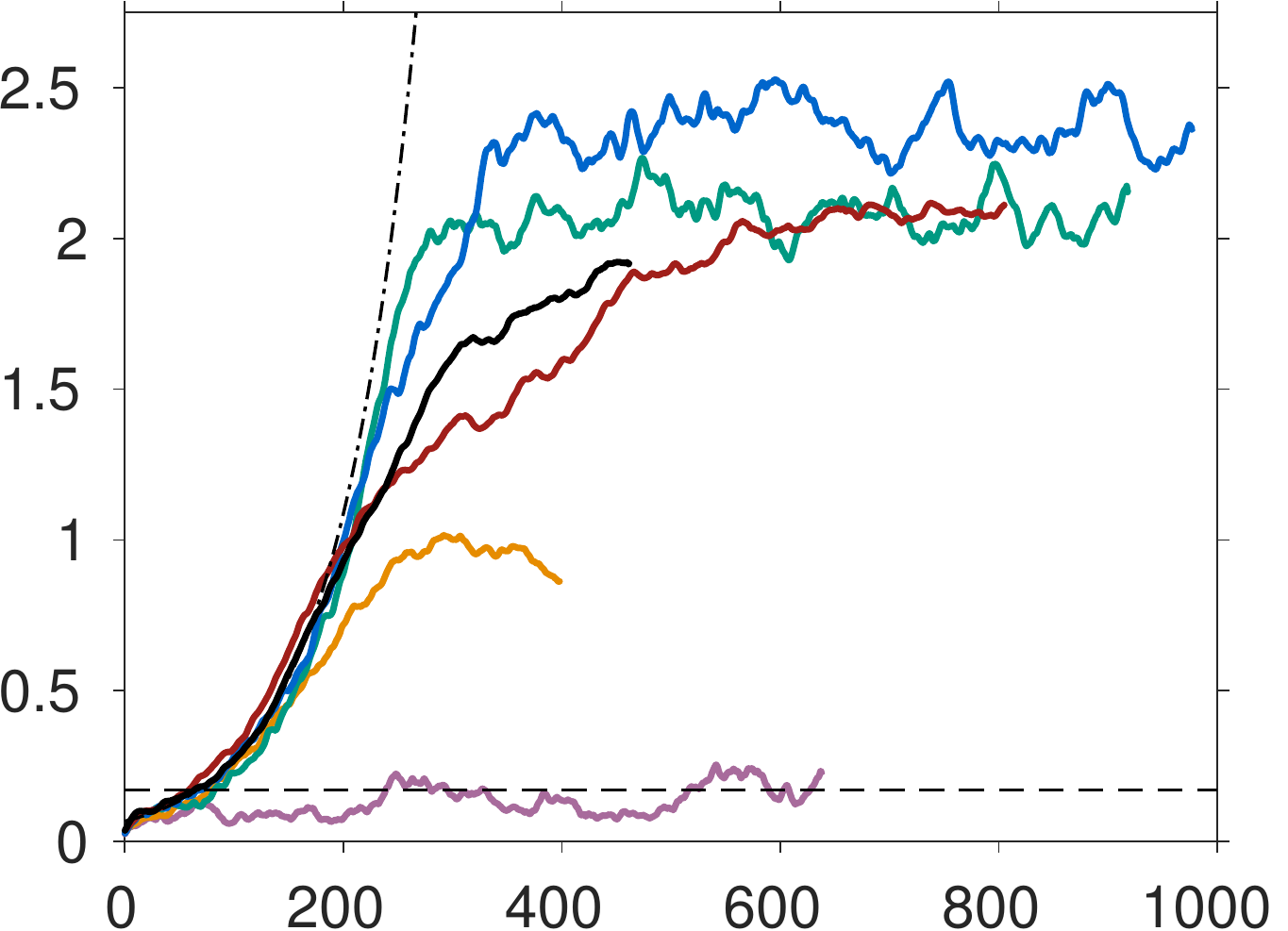}
           \centerline{\small  $t/\tbulk$}
         \end{minipage}\\
         \begin{minipage}{3ex}
           \rotatebox{90}
           {\small \hspace{2ex}$\rmsamplitude/D$}
         \end{minipage}
         \begin{minipage}{.45\linewidth}
           \centerline{(\textit{c})}
           \includegraphics[width=\linewidth]
           {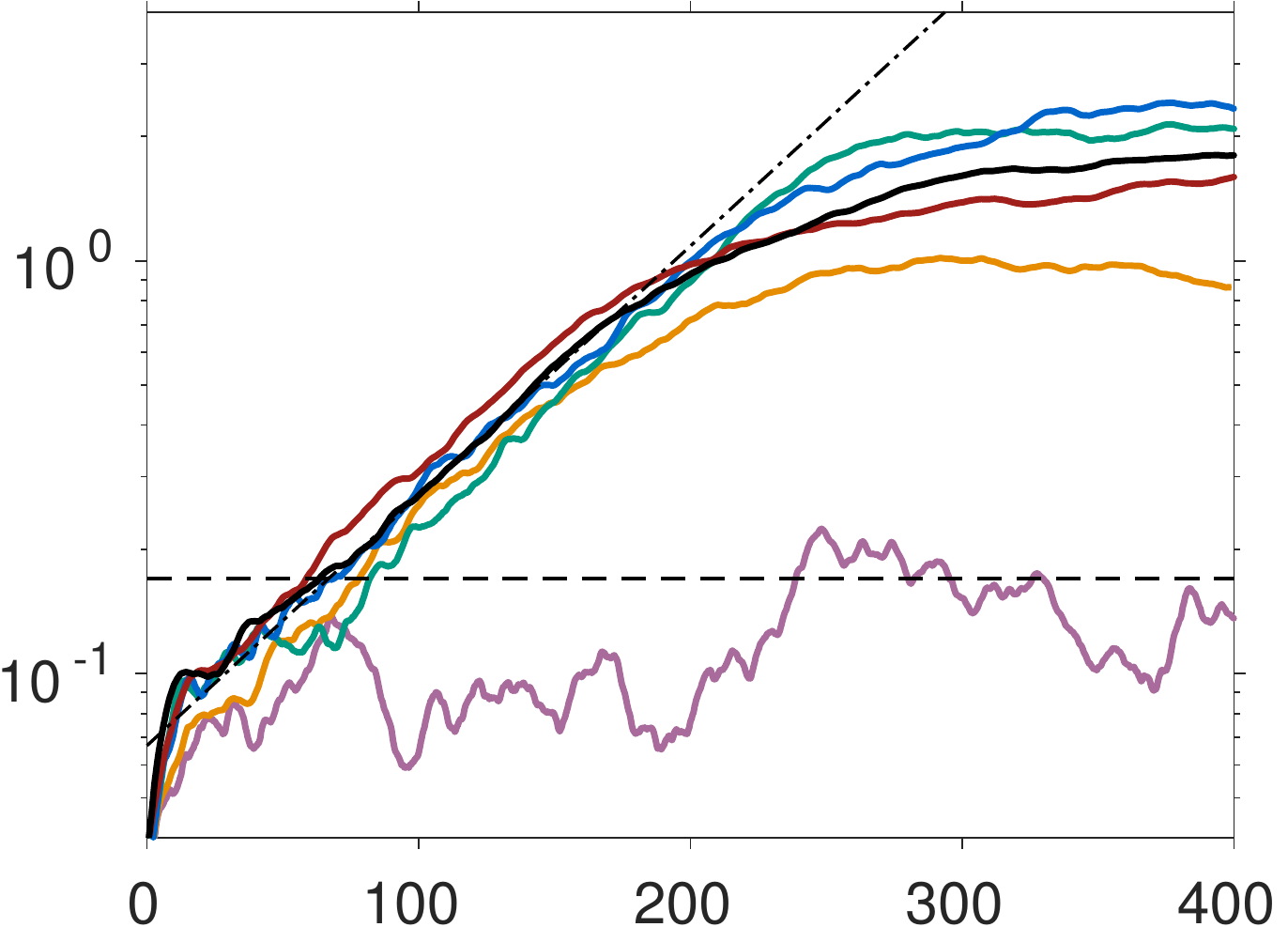}
           \centerline{\small  $t/\tbulk$}
         \end{minipage}
        \caption{%
         (\textit{a})
         Time evolution of the r.m.s.\ sediment bed  height 
         fluctuation normalized by the particle diameter.
         {\color{mypurple}\asolidthick},~case~\caseA;
          {\color{myorange}\asolidthick},~case~\caseB;
         {\color{mygreen}\asolidthick}, case~\caseC;
         {\color{myblue}\asolidthick}, case~\caseD;
         {\color{myred}\asolidthick}, case~\caseE;
         {\color{black}\asolidthick}, case~\caseF.
          The horizontal dashed
          line indicates the value $\rmsamplitude/D\approx0.17$
          which corresponds to the random bed fluctuations of featureless
         flat bed \citep{Coleman2009}.
         (\textit{b}) Same as in (\textit{a}) 
         but the data for cases \caseB\ and \caseE\
         is ensemble averaged over the number of simulations
          performed.
          The dashed-dot line corresponds to an exponential fit curve 
         $\rmsamplitude/D = 0.0668\exp{(0.0140t/\tbulk)}$
         obtained from an average amplitude growth in the initial exponential 
         interval.
        (\textit{c}) Same as in (\textit{b}) 
          but plotted in log-linear scale
         to highlight the exponential growth regime during
           the  first instant of the bed instability.   
         }
         \label{fig:time-evolution-amplitude-all} 
\end{figure}
Figure~\ref{fig:time-evolution-amplitude-all}
shows the time evolution of \rmsamplitude\ for all the considered cases
listed in table~\ref{tab:physical-parameters}.
It is seen that, starting from time $t\ubulk/\hmean\approx 20$,
the value of \rmsamplitude\
\revision{monotonically}{} grows with time 
for all the cases with domain length $\Lx \geq 102.4D$ ($\Lx \geq 4\hmean$),
indicating that the chosen streamwise box dimension of these cases is 
larger than the lower threshold of unstable wavelengths, such
that the box can accommodate
at least one of the unstable modes. 
On the contrary, the evolution
of \rmsamplitude\ in case \caseA\ exhibits no
growth with time except for small fluctuations;  \rmsamplitude\ in
this case is effectively 
bounded by the threshold value $\rmsamplitude/D\approx0.17$ as per
\citet{Coleman2009}. 
This indicates that the streamwise box length 
$\Lx = 76.8D$ ($\Lx = 3\hmean$) is not sufficient to accommodate 
the lower threshold of unstable
modes. Thus, it can be concluded that a cutoff length scale for 
pattern formation (for the considered parameter point) lies in the 
range
$77D < \minimumlambda < 102D$, or in terms of the mean fluid height,
$3\hmean < \minimumlambda < 4\hmean$. Additional simulations
with \Lx\ in the range $3\hmean$ to $4\hmean$ would be required in order to
determine \minimumlambda\ more accurately. 

Furthermore, 
figure~\ref{fig:time-evolution-amplitude-all} highlights the fact 
that  the evolution of the pattern amplitude exhibits 
distinct  regimes of growth.
During the first approximately 150 to 180 bulk time units
(excluding the first of approximately 20 
bulk time units during which the sediment bed initially dilates
after the start of the simulations),   
\rmsamplitude\ is observed to evolve exponentially for all the cases,
with a growth rate which seems to be independent of the chosen domain size.
Note that the time evolution of \rmsamplitude\ for cases \caseB\ and \caseE\ 
presented in figures~\ref{fig:time-evolution-amplitude-all}(\textit{b,c})
is an ensemble 
average over the three separate simulations of the respective cases.  
An exponential curve of the form 
\begin{equation}\label{eq:exponential-fit}
\rmsamplitude/D = A\exp{(Bt/T_b)}
\end{equation}
with $A=0.0668$ and $B=0.0140$,
best fits the average growth over all cases in
this interval. 
In the subsequent growth regime, the trend of the 
amplitude evolution is observed to be markedly different among the 
different cases, 
although all were carried out at the same imposed parameter values.
After the initial exponential growth (up to $t\approx 200\tbulk$), 
and after a small transition
interval of approximately $100\tbulk$, cases
\caseB, \caseC\ and \caseD\ exhibit a
plateau of the pattern amplitude, showing no further growth with
time. The attained final values
are $\rmsamplitude/D = 0.93$ , $2.08$ and $2.36$ for each of these cases,
respectively. The evolution of \rmsamplitude\ for case \caseE\
also attains a plateau regime with a final value of \rmsamplitude\
comparable to that of case \caseC.
However, it settles to this value not immediately after the exponential 
regime, rather it gradually increases (approximately linearly) in the interval
between $t=200\tbulk$ and $600\tbulk$.
On the other hand, since the observation interval of case~\caseF\
is shorter, it has only  covered only the exponential growth
regime by  the end of the simulation. 

The different trend of the amplitude growth observed is closely related 
to the influence of the 
limited computational box size on the initially 
accommodated mean wavelength and its  subsequent evolution. 
In  cases
\caseB, \caseC, \caseD\ and \caseE, even though the
streamwise length of the computational box is larger than the threshold for
pattern formation, due to the fact that $\Lx/\minimumlambda=\mathcal{O}(1)$,
the discrete wavelengths from which the system has to select are very sparse,
and it is not guaranteed that the initially selected wavelength will
be the same as the one which an infinitely long system would select. 
On the other hand, case~\caseF\ has a streamwise box length
\Lx\ which is between $12$ to $16$ multiples of the minimum unstable
wavelength.
Thus the initial
wavelength selected in case~\caseF\ is expected to be sufficiently
close to that of an infinitely long system.
\begin{figure}[t]
   \centering
        \begin{minipage}{2ex}
          \rotatebox{90}
          {\small \hspace{6ex}$\meanlambda/D$}
        \end{minipage}
        \begin{minipage}{.45\linewidth}
          \centerline{(\textit{a})}
          \includegraphics[width=\linewidth]
          {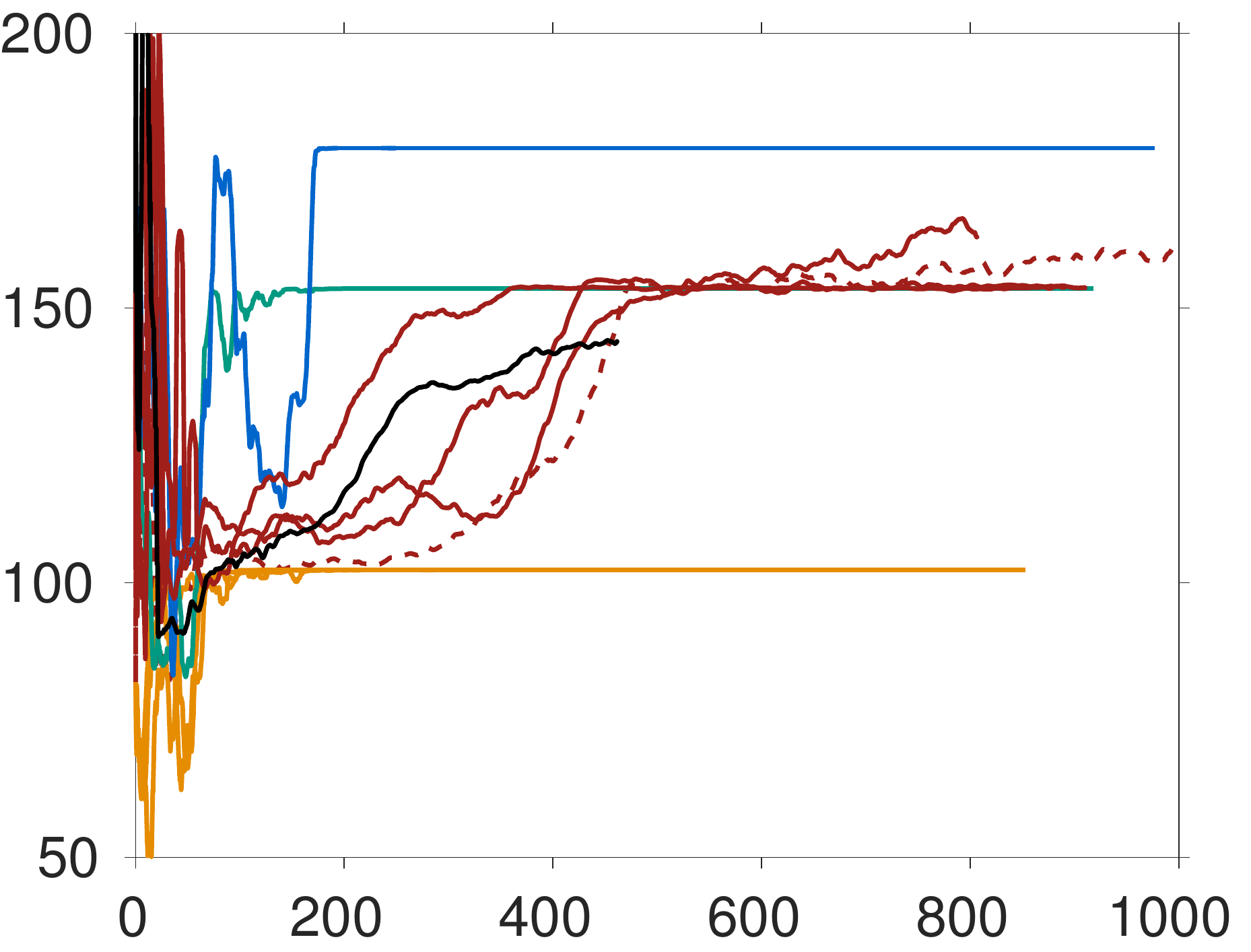}
          \centerline{\small  $t/\tbulk$}
        \end{minipage}
        \hfill
        \begin{minipage}{2ex}
          \rotatebox{90}
          {\small \hspace{6ex}$\meanlambda/D$}
        \end{minipage}
        \begin{minipage}{.45\linewidth}
          \centerline{(\textit{b})}
          \includegraphics[width=\linewidth]
          {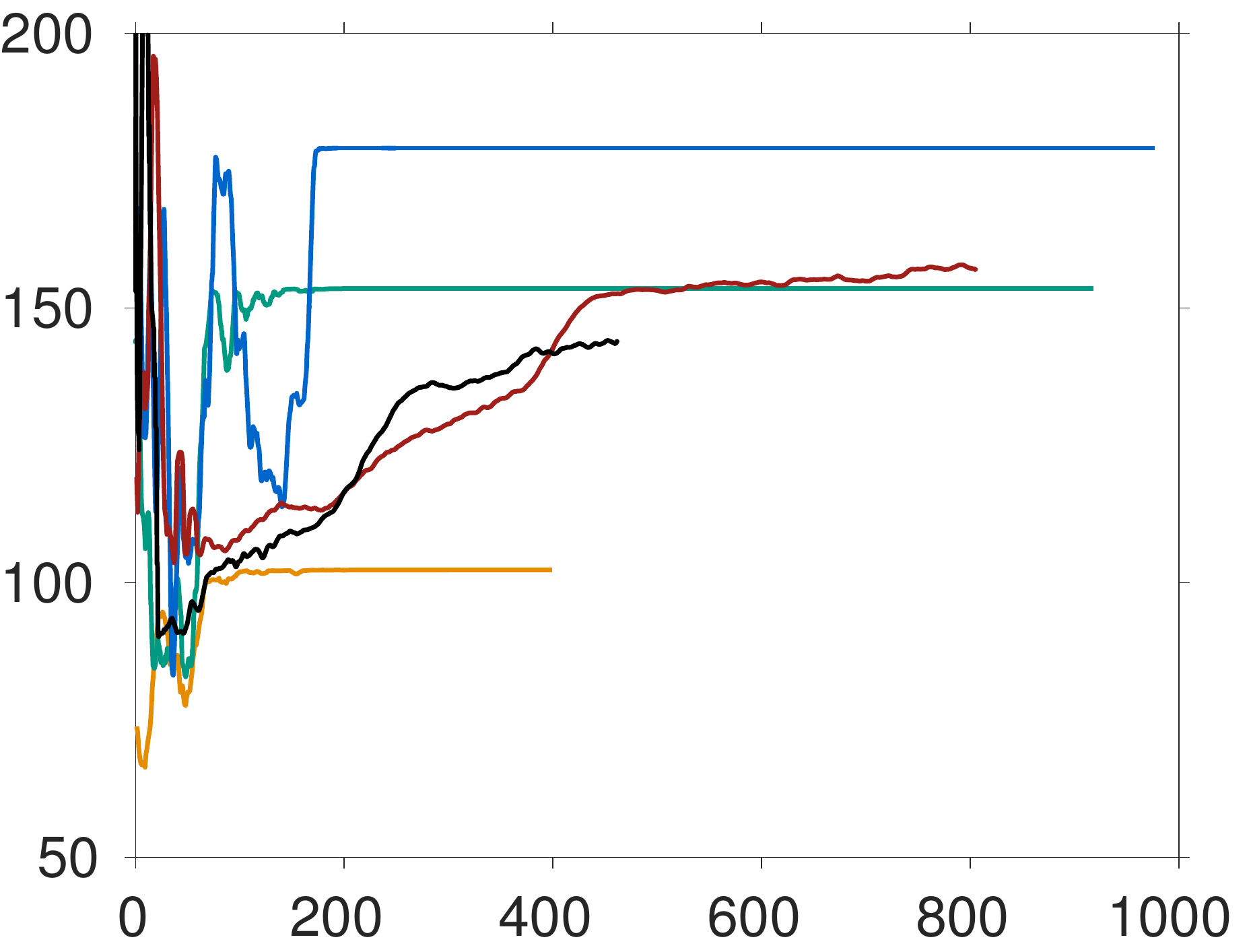}
          \centerline{\small  $t/\tbulk$}
        \end{minipage}
        \caption{%
         (\textit{a})
         Time evolution of the mean wavelength \meanlambda\ of the sediment bed
         height normalized by the particle diameter. 
       Color coding similar as in 
       figure~\ref{fig:time-evolution-amplitude-all}
       (\textit{b}) the same as in (\textit{a}) but the data for 
        cases \caseB\ and \caseE\
         is ensemble averaged over the number of simulations performed.}      
         \label{fig:time-evolution-wavelength-all} 
\end{figure}
Figure~\ref{fig:time-evolution-wavelength-all} shows the time
evolution of the mean pattern wavelength. In close correlation with
the evolution of the amplitude, one can see that the adopted
computational box size strongly influences the selected mean wavelength and
its subsequent evolution. In cases \caseB\ and \caseC, 
\meanlambda\ jumps within the first 100 to 200 bulk time units to the 
maximum possible wavelength, i.e.\ $\meanlambda=L_x$, and evolves 
constrained to this value. 
This indicates that in these boxes, the 
system is forced to select the most unstable wavelength from the 
available modes, which turns out to be
$\meanlambda\approx 102D$ for case \caseB\ and
$\meanlambda\approx 154D$ for case \caseC. The value of the mean
wavelength of case \caseD\ is observed to oscillate between the harmonics
$\lambda_1=L_x$ and $\lambda_2=L_x/2$ and finally settles at
 $\meanlambda = \lambda_2\approx 179D$ at $t/\tbulk\approx 180$.
On the other hand, 
from the ensemble averaged evolution of wavelength for case \caseE,
it can be seen that the system initially selects a wavelength
$\meanlambda\approx \lambda_3 \approx 102D$, which is very close to the
wavelength selected by the large box size case \caseF. It then grows
monotonically and settles at a wavelength 
$\meanlambda \approx \lambda_2 \approx 154D$ at approximately
$t=500\tbulk$, subsequently evolving constrained close to this wavelength until
the end of the simulation interval. Again, the constraint of the
evolution of the wavelength is an indication of
the influence of box size on the subsequent evolution processes,
although the box could be considered marginally sufficient to capture
the initial wavelength.
The non-linear nature of the evolution process is also highlighted by
observing the difference in the evolution of the wavelength for
the three independent simulations of case \caseE\ (which only differ
from one another by the respective initial condition) 
in the first interval up to $t\approx 500\tbulk$. 
In case \caseF, the availability of relatively fine graded harmonics 
allows for a more steady growth of the mean wavelength. In the 
initial time $t\approx 100\tbulk$ up to $t\approx 200\tbulk$, 
the selected wavelength grows from  
$\meanlambda\approx \lambda_{12}=L_x/12$ to
$\meanlambda\approx \lambda_{11}=L_x/11$. It exhibits a further growth
and  attains a value 
\revision{$\meanlambda\approx 125D$}{%
          $\meanlambda\approx 145D$}
at the end of the 
simulation interval.

The above discussed evolution of the amplitude and 
mean wavelength are integral representations of
the individual modes which 
make up the resolved discrete spectrum. On the other hand, from
stability analysis point of view, it might be more interesting to
analyze the time-evolution of the individual modes, since
for small amplitudes these can be expected to grow independently.
To this end,
we have computed the instantaneous 
\revision{%
  single-sided amplitude spectrum $A_j$ which is defined as
  [deleted equation] %
  where $\hat{h}_{bj}$ is the $j$th harmonic of the 
  Discrete Fourier Transform of $\hbed^\prime$. }
{%
  single-sided amplitude spectrum $A_j(t)$ (for non-negative
  wavenumbers $j\geq0$)  
  which is defined as twice the absolute value of the coefficient
  $\hat{h}_{bj}$, 
  where $\hat{h}_{bj}(t)$ is the $j$th harmonic of the 
  Discrete Fourier Transform of the bed-height perturbation
  $\hbed^\prime(x,t)$. 
  For future reference let us denote with $S_{h_bh_b}(\kappa_j)$ 
  the power spectral-density corresponding to the $j$th Fourier mode.
}
In order to provide a dispersion relation, first we assume that,
in the initial exponential growth regime of the r.m.s. bed height
fluctuation (cf.\ figure~\ref{fig:time-evolution-amplitude-all}),
the  individual modes exhibit an exponential growth as well. Strictly speaking,
this assumption is only true when the bed fluctuations are so small that
linear instability holds. 
Nevertheless, it can be safely assumed
that the non-linear interaction among the modes is weak in the first 
approximately $150$--$180$ bulk time units
\revision{}{(in the complementary hypothetical experiment, this time 
duration amounts to $7$--$9$ seconds)}. 
Following, we fit an exponential curve of the form
\begin{equation}\label{eq:amplitude-growthrate}
A_j(t) = A_{j0}\exp{(\sigma_jt)}
\end{equation}
to the time evolution of the amplitude of each mode in the interval 
between $t\approx 20\tbulk$ up to $t\approx 180\tbulk$ 
and infer the
growth rate $\sigma_j(\kappa_j)$,
where $\kappa_j \equiv 2\pi j/L_x$ is the wavenumber
of the $j$th harmonic.
\begin{figure}
   \centering
        \begin{minipage}{2ex}
          \rotatebox{90}
          {\small \hspace{6ex}$\sigma_j\tbulk$}
        \end{minipage}
        \begin{minipage}{.45\linewidth}
          \includegraphics[width=\linewidth]
          {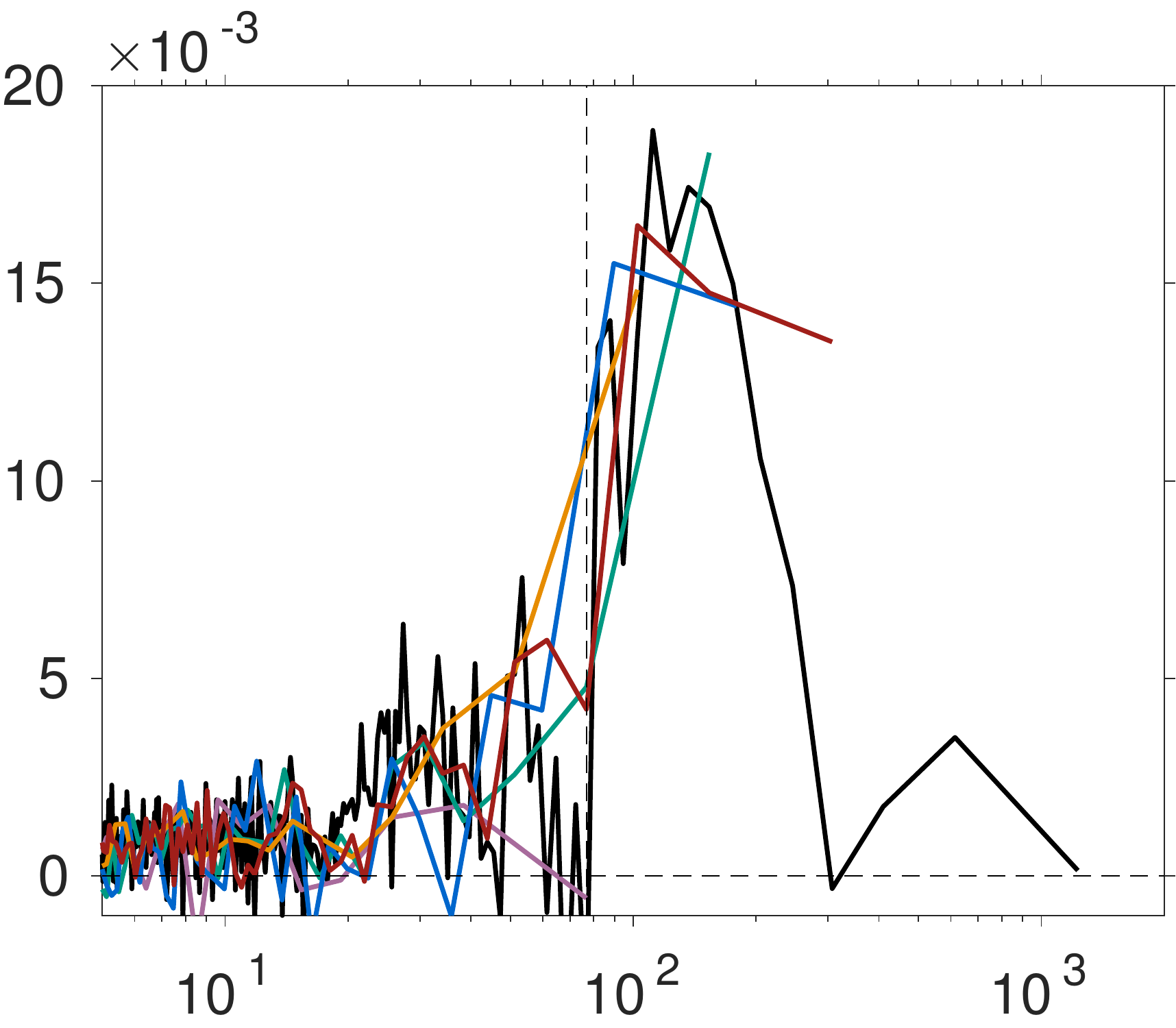}
          \centerline{\small  $\lambda_jD$}
        \end{minipage}
        \caption{%
         \revision{Dispersion relation:}{}
         Amplitude growth rate $\sigma_j$ (as per 
         definition \ref{eq:amplitude-growthrate}) 
         as a function of the wavelength in the
         first exponential growth regime.
         {\color{mypurple}\asolidthick},~case~\caseA;
          {\color{myorange}\asolidthick},~case~\caseB;
         {\color{mygreen}\asolidthick}, case~\caseC;
         {\color{myblue}\asolidthick}, case~\caseD;
         {\color{myred}\asolidthick}, case~\caseE;
         {\color{black}\asolidthick}, case~\caseF.
         The vertical dashed line represents the value of the streamwise
         box length in case~\caseA\ which does not feature patterns.
         }      
         \label{fig:dispersion-relation-all} 
\end{figure}

\begin{figure}[t]
   \centering
        \begin{minipage}{2ex}
          \rotatebox{90}
          {$(\hbedmeanphase - \min{(\hbedmeanphase)})/\meanlambda$}
        \end{minipage}
        \begin{minipage}{.45\linewidth}
          \centerline{(\textit{a})}
          \includegraphics[width=\linewidth]
          {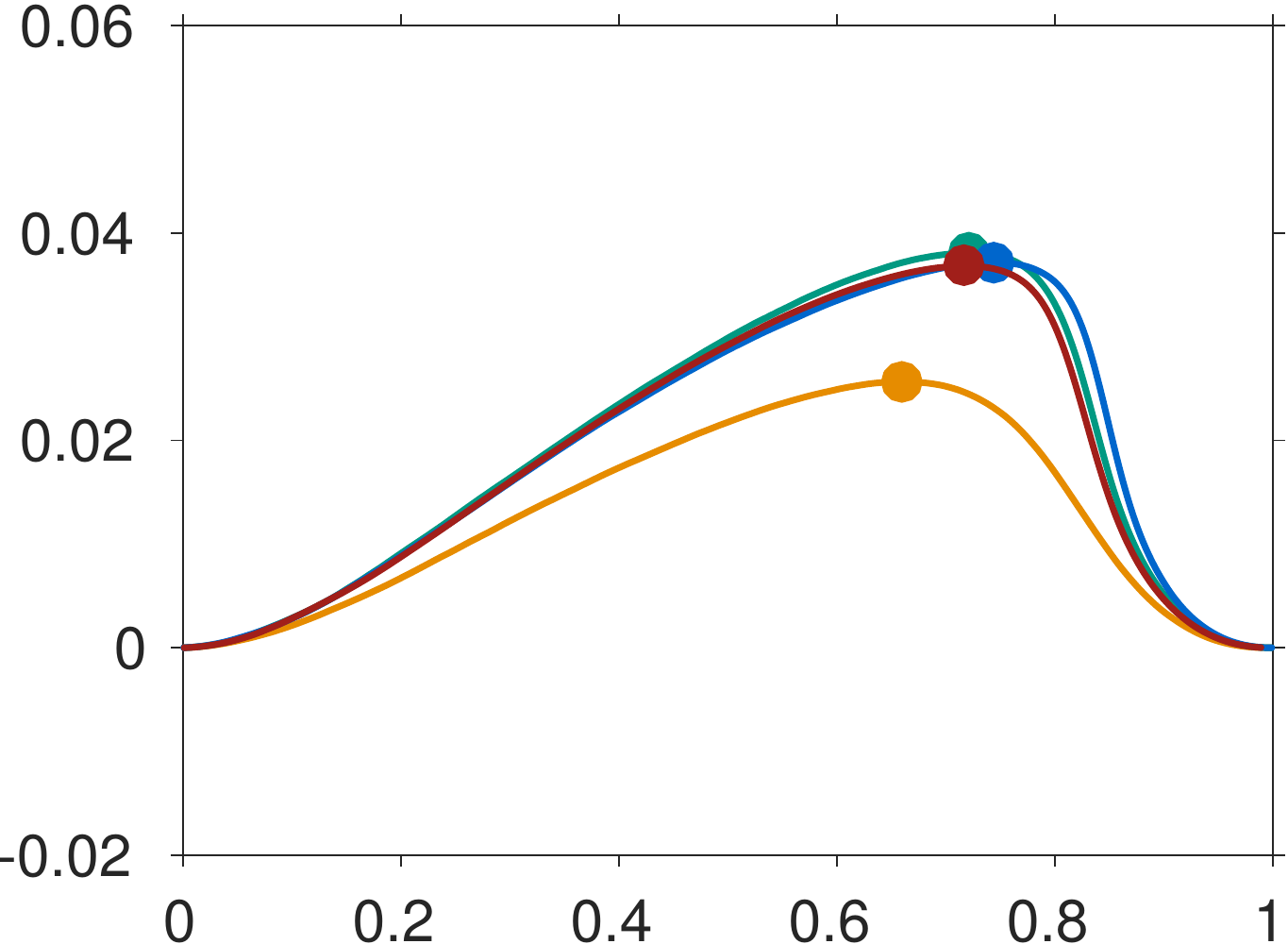}
          \centerline{$\xtilde/\meanlambda$}
        \end{minipage}
        \hfill
        \begin{minipage}{2ex}
          \rotatebox{90}
          {$\langle S_{h_bh_b}\rangle_t/D^2$}
        \end{minipage}
        \begin{minipage}{.45\linewidth}
          \centerline{(\textit{b})}
          \includegraphics[width=\linewidth]
          {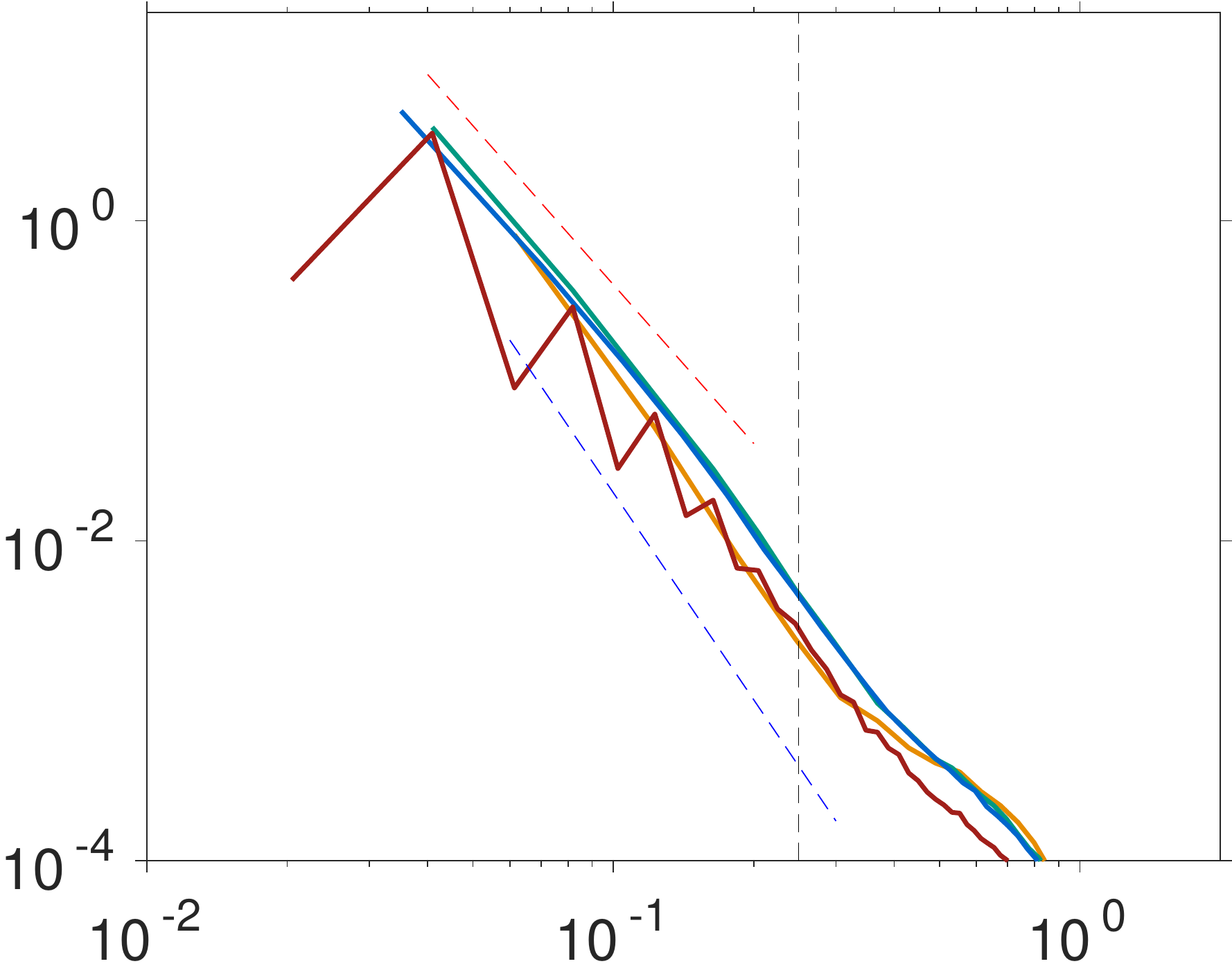}
          \centerline{$\kappa_jD$}
        \end{minipage}
        \caption{%
          (\textit{a})
         Mean two-dimensional profile of the patterns averaged over the 
         last steady \revision{dune}{ripple} propagation interval, and over the number of
         \revision{dunes}{ripples} when applicable, of each case.
         The location of the maximum of each profile is shown by the
         filled symbols. Note that the profiles are plotted not to
         scale and thus the aspect ratio is visually exaggerated.
         (\textit{b}) The corresponding
         spectra of the sediment bed height fluctuation.
          In both plots:
          {\color{myorange}\asolidthick},~case~\caseB;
         {\color{mygreen}\asolidthick}, case~\caseC;
         {\color{myblue}\asolidthick}, case~\caseD;
         {\color{myred}\asolidthick}, case~\caseE.
         Power \revision{curves}{laws} 
          with exponent \revision{$-3.5$}{$-3.3$} 
         and $-4.3$ are shown by the red and blue dashed lines
         respectively.
         The vertical dashed-line represents the value of the 
         average fluid height $\hmean\approx 25.27 D$ (averaged over 
         all cases).
         }      
         \label{fig:dune-geometry-vs-wavelength}
\end{figure}
Figure~\ref{fig:dispersion-relation-all} shows the growth
rate
as a function of the wavelength $\lambda_j\equiv 2\pi/\kappa_j$.
As is expected, 
\revision{none of the modes are seen to grow for}{%
none of the modes are seen to grow at any significant rate in}
 case \caseA.
Moreover, consistent with the evolution of the
mean wavelength, the fastest growing mode in cases \caseB\ and \caseC\
is $\lambda_1=L_x$, with a growth rate which is comparable to that
of the r.m.s. bed height \rmsamplitude\ (coefficient $B$ in 
equation \ref{eq:exponential-fit}). The next few modes in these two
cases also exhibit 
a small but noticeable growth rate although their wavelength is 
smaller than the threshold. This could be attributed to 
non-linear effects. A similar trend is observable for cases
\caseD\ and \caseE, except that in these cases, it is not only one 
harmonic which is growing the fastest. In case \caseD\ both 
$\lambda_1=L_x$ and  $\lambda_2=L_x/2$ grow at a comparable rate,
indicating that both harmonics contribute substantially to the growth of 
\rmsamplitude. On the other hand, in case \caseE, in addition to
the 
expected fastest growing modes $\lambda_3=L_x/3$ and
$\lambda_2=L_x/2$, the first harmonic $\lambda_1=L_x$ is also seen to
contribute to the overall growth. 
It is not guaranteed that the box length of \caseF\ is large
enough to capture  all
the unstable modes (i.e.\ the upper threshold of pattern formation).
Nevertheless,
the fact that $\sigma_j$ exhibits  
a clear local maximum is an indication that the box is sufficient to 
capture the most amplified mode(s) which turn out to be in the range
$\lambda_8$ to $\lambda_{12}$.

Finally, it is worth mentioning that the computational box sizes are
chosen such that 
a turbulent flow state is sustained.
This is confirmed by monitoring
the time evolution of the box-averaged turbulent kinetic energy (plots
not shown). Note that, in smooth wall channel flows, the minimum
spanwise and streamwise box dimensions required in order for
turbulence to be self-sustained, are 
$\Lz^+ \approx 100$ and $\Lx^+ \approx 350$, 
respectively \citep{Jimenez1991}.  The minimum box
dimensions adopted in our study (corresponding to case~\caseA) are
substantially larger, i.e.\ $\Lx^+ = \Lz^+\approx 740$.
\subsection{\revision{Dune}{Ripple} morphology}
\label{subsec:dune-morphology}
An additional important advantage of adopting a relatively `small'
computational box length in cases \caseB, \caseC\ and \caseD\ is that,
once the maximum possible wavelength is attained, further ripple
dimension growth is effectively hindered. Thus, the accommodated
ripple subsequently evolves steadily maintaining its
dimensions and propagation velocity.  Similarly, in case \caseE,
within the simulated interval, the ripples evolve steadily constrained
approximately at a mean wavelength $\meanlambda=L_x/2$.  Such a numerical
manipulation, which experimentally is not possible, allows us to
address steady state 
\revision{dune}{bedform} 
characteristics at a particular
wavelength. In the following, we analyze the two-dimensional shape and
the propagation velocity of the \revision{dunes}{ripples}.

It is well know that bedforms of finite amplitude evolve towards an
asymmetric shape even in the absence of flow separation
\citep{Best2005,Colombini2008a,Seminara2010}. For instance,
spatio-temporal plots provided by \citet{Ouriemi2009b} show that the
evolution of the bedforms towards their asymmetric shape is a
phenomenon of both the laminar `small dunes' which exhibit no flow
separation in their leeside as well as the turbulent `vortex dunes'
which are characterized by flow separation downstream of their
crests. The degree of asymmetry is generally larger in the latter
as a result of the flow recirculation.
Moreover, the animations of particle motion available online at
\href{http://dx.doi.org/10.1017/jfm.2014.284}
     {http://dx.doi.org/10.1017/jfm.2014.284} 
and the space-time plots depicted in figure 4 of
\citet{Kidanemariam2014} have qualitatively shown the asymmetric
evolution of sediment bed shape in good agreement with what is
observed experimentally.

In order to quantitatively characterize the statistically 
two-dimensional shape of
the 
\revision{dunes}{bedforms}, 
in the steady 
\revision{dune}{ripple} 
evolution interval of cases \caseB, 
\caseC, \caseD\ and \caseE,  we have evaluated the phase-averaged 
fluid-bed interface which is defined as follows:
\begin{equation}\label{eq:phase-averaged-interface}
\hbedmeanphase(\xtilde) = \langle \hbed (\xtilde,t)\rangle_t\;,
\end{equation}
where $\xtilde\equiv  x-\celerity t$ is the $x$-coordinate in a frame of
reference which is moving at the mean 
\revision{dune}{ripple} 
propagation 
velocity \celerity\ (see  section~\ref{subsec:propagation-speed} for the
definition of \celerity). Note that, in cases \caseB, \caseC\ and \caseD, 
\hbedmeanphase\ corresponds to the mean profile of a single 
\revision{dune}{ripple} 
while in case \caseE, \hbedmeanphase\ is an average profile over the two
available \revision{dunes}{ripples}.  

As shown Figure~\ref{fig:dune-geometry-vs-wavelength}(\textit{a}),
it can be seen that in all cases, the 
\revision{dune}{ripple} 
profile is clearly asymmetric with 
an upstream face (upstream of the crest) approximately three times
longer in streamwise length and thus milder in bed slope than the 
downstream face. 
To within the statistical uncertainty, the bed shape
of cases \caseC, \caseD\ and \caseE, once scaled by the corresponding
mean wavelength, seems to fairly collapse into a single shape, exhibiting
small differences.
The attained value of the aspect ratio, which is defined as the
ratio between the height of the 
 \revision{dune}{ripple} 
$H_D = \max{(\hbedmeanphase)} - \min{(\hbedmeanphase)}$, (i.e.\
the wall-normal distance between the trough and the crest) 
and the mean wavelength \meanlambda, is
approximately 0.037. Moreover, the 
degree of asymmetry, which can be defined as
the ratio between the streamwise distance from the crest to the
downstream trough and the mean wavelength, 
is approximately 0.28. On the other hand, the mean 
 \revision{dune}{ripple} 
shape of case \caseB\ is seen to exhibit a noticeable smaller value
of the aspect
ratio (approximately 0.026) and a larger value
of the degree of asymmetry 
(approximately 0.35) than the 
other cases. This difference is an indication of the 
\revision{dune}{ripple} 
 shape evolution
process. Some experimental studies reported that, 
during the natural ripple coarsening process, the aspect 
ratio of the 
\revision{dune}{ripple} 
 geometry  evolves with time and attains 
a value of approximately 1/15 ($0.067$) once the  
steady state  shape is attained \citep{Charru2013a}. 
On the other hand, \citet{Fourriere2010}, based on
field measurements, report a steady state aspect ratio of approximately
$0.045$.
In our configurations, since the final ``natural'' steady-state regime
is not yet reached, we can not confidently say that the 
attained values of the aspect ratios are representative of those values which
would naturally be attained. Nevertheless, the fact that the shape of
the 
\revision{dune}{ripple} 
 with mean wavelength $\meanlambda\approx 150D$ and 
that with $\meanlambda\approx 180D$ do not exhibit substantial 
differences could be an indication that the 
\revision{dune}{ripple} 
 shape is not far
from its final invariant shape. This aspect requires further investigation.

It is generally believed that the non-linear interaction between the
evolving sediment bed and the driving flow results in the above
mentioned asymmetric 
\revision{dune}{ripple} 
 shape \citep{Charru2013a}.
That is, although the 
\revision{dunes}{ripples} 
have one well-defined length scale,
(the mean wavelength \meanlambda\ defined in 
equation~\ref{eq:mean-wavelength}), 
a wide spectrum of modes  
interact non-linearly to maintain a shape-invariant 
\revision{dune}{ripple} 
 which 
propagates downstream.  
There are several theoretical
studies as well as field 
and experimental measurements which report that the spectra
of fully developed bedforms exhibit a $-3$ power law relationship
with respect to the large wavenumbers,
bounded by a threshold value 
\citep{Hino1968,Jain1974,Nikora1997b,Coleman2011}.
To gain further insight,
we present in figure~\ref{fig:dune-geometry-vs-wavelength}(\textit{b}) 
the spectra $\langle S_{h_bh_b}\rangle_t$ of the sediment bed height 
fluctuation averaged over the
steady \revision{dune}{ripple} evolution interval of the above mentioned cases.
It can be seen that, for the dominant modes (with $\lambda_j \gtrsim \hmean$),
$\langle S_{h_bh_b}\rangle_t$
of all the cases (only for the even modes of case \caseE) features a power 
law variation as a function of $\kappa_j$. The value of the scaling 
exponent in cases \caseC,
\caseD\  and for the even modes in \caseE\  is observed to be approximately 
\revision{$-3.5$}{$-3.3$} 
whereas that of case \caseB\ is approximately $-4.3$. The fact that case
\caseB\ has a larger value of the exponent than that of the other cases is
again an indication of the evolution process.  
That is, the
influence  on the mean \revision{dune}{ripple} shape of the modes with a smaller 
wavelength than the dominant one, is relatively smaller 
compared to the other cases.  Moreover, the fact that we have not
attained a naturally developed mature \revision{dune}{ripple} state could be a reason why the
attained exponents are \revision{}{slightly}
larger than the $-3$ value reported in the
literature. The reason for the odd-even separation of the spectra in 
case \caseE\ could be explained by fact that the domain length accommodates
two ripples and thus modes with wavelength $L_x/(2i)$ mainly contribute to
the attained shape. 
\subsection{\revision{Dune}{Ripple} migration velocity}
\label{subsec:propagation-speed}
 The migration velocity of bedforms is an important quantity of
 interest in the study of morphodynamics. For instance, the rate at
 which bedforms propagate is closely related to the rate of sediment
 transport \citep{Coleman1994,Nikora1997b,Betat2002a,
 Coleman2009a,Ouriemi2009b,Seminara2010,Coleman2003}. 
From the sediment mass conservation equation (Exner equation), 
it can be easily shown that, for a statistically shape-invariant 
two-dimensional \revision{dune}{ripple} with height $H_D$ and which migrates
downstream at a constant velocity \celerity, 
\begin{equation}\label{eq:exner-qp-vs-dune-celerity}
\partflowrate -\langle q_{p,min}\rangle \approx \beta \phibed \celerity H_D\,,
\end{equation}
where \partflowrate\ is the mean particle flowrate, which is assumed to be
constant, while $\langle q_{p,min}\rangle$ is the corresponding minimum value 
at the trough of the \revision{dune}{ripple},  \phibed\ is the mean solid volume fraction
of the sediment bed and $\beta$ is a \revision{dune}{ripple} shape parameter.
From the approximation (\ref{eq:exner-qp-vs-dune-celerity}), it is evident 
that the \revision{dune}{ripple} migration velocity is
inversely proportional to the \revision{dune}{ripple} height (and thus the \revision{dune}{ripple} mean
wavelength) whereas it is directly proportional to the mean particle 
flow rate \citep{Charru2013a}.
In the following, based on the DNS data, we evaluate the mean \revision{dune}{ripple} 
migration velocity and assess its relation to the mean particle flow rate.

In the aforementioned steady evolution interval,
an average migration velocity  of the patterns \celerity\
can be determined from the 
shift of the maximum of space--time correlation function of the 
fluid--bed interface fluctuation 
$\hbed^\prime$ \citep[][]{Nikora1997b,Coleman2011}. 
First, the space--time correlation function is defined as follows:
\begin{equation}
   R_{ht}(\rsep,\tsep) = 
      \langle\, 
      \hbed^\prime(x,t_1) \cdot
      \hbed^\prime(x+\rsep,t_1+\tsep)\,\rangle_{x}
     \quad
     \forall \tsep \in 
     \left[0,\; t_2 -t_1\right]
\end{equation}
where $t_1$ and $t_2$ are the start and end times of the considered interval. 
The streamwise location $x_{max}$ of the maximum 
of $R_{ht}$, which is given by 
\begin{equation}
r_x^m(\tsep) =  x_{max}(\tsep) \;|\; \forall\, \rsep\in [-\Lx/2,\;\Lx/2] :
              R_{ht}(\rsep,\tsep)\le R_{ht}(x_{max},\tsep)\,,
\end{equation}
varies approximately linearly with respect to $\tsep$, and thus the slope
of the line which is fitted to the data $r_x^m$ versus $\tsep$ gives
an accurate estimate of \celerity.

\begin{figure}[t]
   \centering
        \begin{minipage}{3ex}
          \rotatebox{90}{$\celerity^+$ 
          }
        \end{minipage}
        \begin{minipage}{.45\linewidth}
          \centerline{(\textit{a})}
          \includegraphics[width=\linewidth]
          {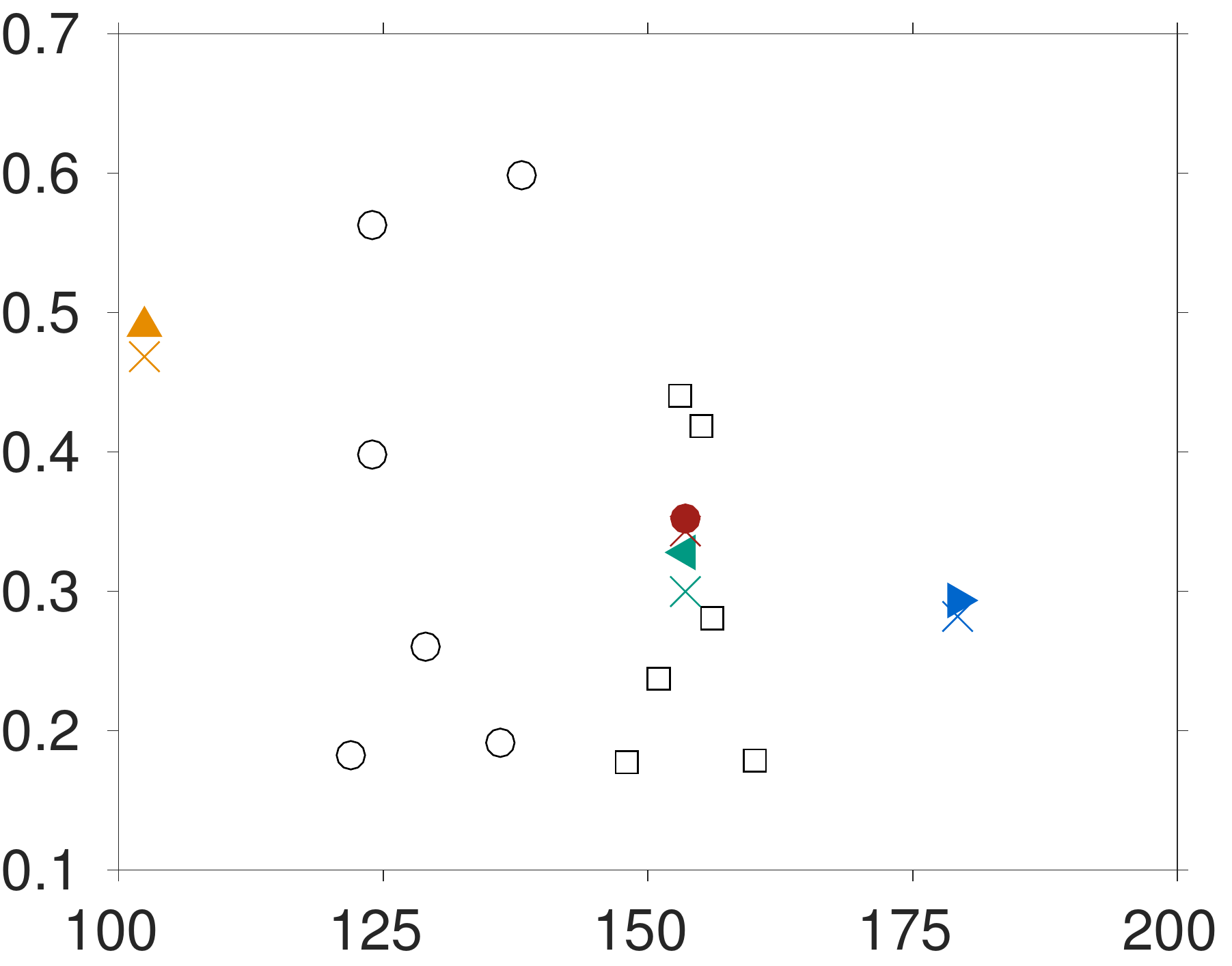}
          \centerline{$\meanlambda/D$}
        \end{minipage}
        \hfill
        \begin{minipage}{2ex}
          \rotatebox{90} {$\langle c_j\rangle^+$}
        \end{minipage}
        \begin{minipage}{.45\linewidth}
          \centerline{(\textit{b})}
          \includegraphics[width=\linewidth]
          {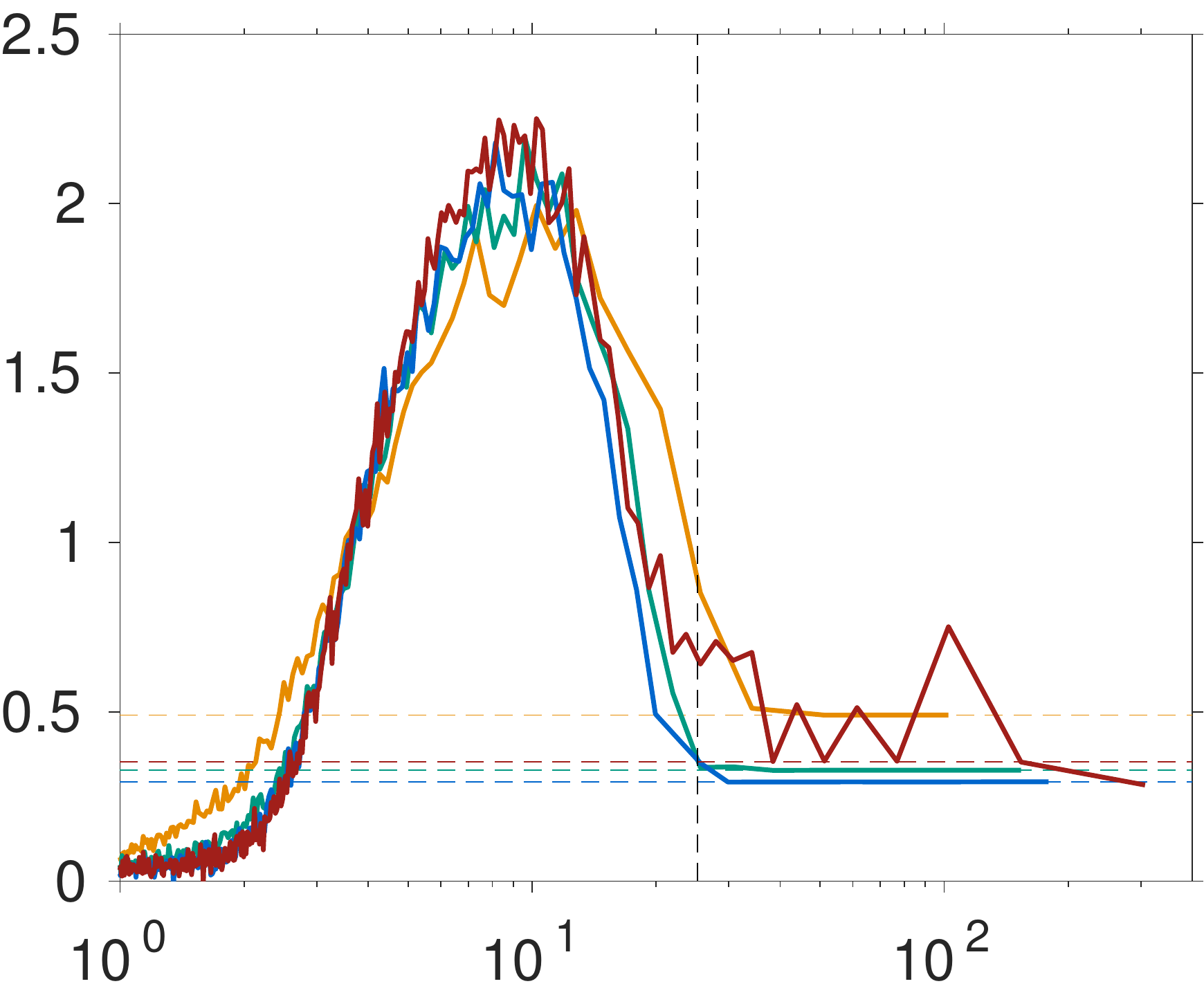}
          \centerline{$\lambda_j/D$}
        \end{minipage}
        \caption{%
         (\textit{a})
         Average propagation speed of the patterns, normalized in 
         wall units, as a function of the  mean wavelength
         \revision{}{(computed in the final steady ripple evolution 
          interval as shown in table~\ref{tab:numerical-parameters})}.
           {\color{myorange}\asolidtriup},~case~\caseB;
         {\color{mygreen}\asolidtrileft}, case~\caseC;
         {\color{myblue}\asolidtriright}, case~\caseD;
         {\color{myred}\asolidcircle}, case~\caseE.
         The corresponding cross marks (\across) represent the 
         mean \revision{dune}{ripple} migration velocity estimated by 
         relation~(\ref{eq:exner-qp-vs-dune-celerity}).
         The data points in the turbulent channel flow experiment 
         of \citet{CardonaFlorez2016} for their particles
         with diamter $D=550\mu {\rm m}$ are shown by the open
         symbols \aopencircle\ (initial phase of ripple evolution) and 
         \aopensquare\ (final phase of the evolution). 
          (\textit{b}) 
          \revision{Phase speed dispersion relation}{%
           Mean propagation speed of individual Fourier modes} 
          in the final steady \revision{dune}{ripple}
          evolution interval.
          {\color{myorange}\asolidthick},~case~\caseB;
         {\color{mygreen}\asolidthick}, case~\caseC;
         {\color{myblue}\asolidthick}, case~\caseD;
         {\color{myred}\asolidthick}, case~\caseE.
         The corresponding mean \revision{dune}{ripple} migration velocity is plotted
         by the horizontal dashed lines.
         The vertical dashed-line represents the value of the 
         average fluid height $\hmean\approx 25.27 D$ (averaged over 
         all cases).
         In both plots, ensemble averaging is performed for cases  
         \caseB\ and \caseE\ over the available number of runs.
         }      
         \label{fig:propagation-speed-vs-wavelength}
\end{figure}
Figure~\ref{fig:propagation-speed-vs-wavelength}\textit{a} 
shows the migration velocity, normalized by \ufric,
as a function of the mean wavelength for cases 
\caseB, \caseC, \caseD\ and \caseE. The variation of
\celerity\ exhibits an inverse-relationship with the \revision{dune}{ripple} wavelength,
varying from a value $\celerity^+\approx 0.49$ for 
case \caseB\ down to a value $\celerity^+\approx 0.3$ for 
case \caseD. It should be noted however that, as will be shown in
section~\ref{subsec:particle-flow-rate}, 
as a consequence of an
increase of the mean bottom shear stress, the value of the mean
particle flow rate increases with increasing mean wavelength. 
Thus the observed trend of the \revision{dune}{ripple} migration velocity is a net
result of an increase of particle flow rate as well as an increase
of \revision{dune}{ripple} dimensions. 
In the following, we assess the mass-conservation-based relationship 
between the mean particle flow rate and the \revision{dune}{ripple} migration velocity, 
by reverse-computing the latter from  relation 
(\ref{eq:exner-qp-vs-dune-celerity}).
To this end, a mean particle flow rate \partflowrate\ is computed
over the steady \revision{dune}{ripple} migration interval.
(cf.\ section \ref{subsec:particle-flow-rate}).
  The average minimum 
particle flow rate at the \revision{dune}{ripple} troughs $\langle q_{p,min}\rangle$ 
was computed as well. It turns out that the value 
$\langle q_{p,min}\rangle$ is 
negligibly small in cases \caseC, \caseD\ and \caseE, whereas in case~\caseB,
it attains a value of $\langle q_{p,min}\rangle \approx 0.2\partflowrate$. 
Furthermore, the \revision{dune}{ripple} shape parameter $\beta$ is defined as 
the ratio between the two-dimensional area of the \revision{dune}{ripple} to that of a 
bounding rectangle of dimensions $H_D$ and
$\meanlambda$, defined as 
\begin{equation}\label{eq:dune-shape-parameter}
\beta = \frac{1}{H_D\meanlambda}\int_{0}^{\meanlambda}
        (\hbedmeanphase(x)-\min{(\hbedmeanphase)}){\rm d}x\,.
\end{equation}
In most of the engineering-purpose bedload transport rate estimators,
the value of $\beta$ is typically taken to be $0.5$, assuming that the 
shape of the \revision{dune}{ripple} is triangular, while some adopt values up to
$\beta=0.6$ based on field and experimental measurements
(see e.g.\ \citet{Gaeuman2007} and
references therein). For the mean \revision{dune}{ripple} geometries encountered in the 
present study, a value of $\beta$ in the range between $0.5$ and $0.52$
is recovered after 
evaluating equation (\ref{eq:dune-shape-parameter}) for all the cases.
Subsequently, the mean \revision{dune}{ripple} migration velocity is estimated based on 
relation (\ref{eq:exner-qp-vs-dune-celerity}) and the result is 
presented in
figure~\ref{fig:propagation-speed-vs-wavelength}\textit{a}.
It can be seen that the estimation yields values which are 
in good agreement with the actual computed \revision{dune}{ripple} migration velocity values. 

Additionally, figure~\ref{fig:propagation-speed-vs-wavelength}\textit{a}
shows recent experimental measurements
of ripple migration velocity in a closed-conduit setup \citep{CardonaFlorez2016}.
The experimental data shown corresponds to particles with diameter
$D=550\mu {\rm m}$ 
(\revision{$\Reb\approx 9000$, $\dratio=2.5$, $\hmean/D\approx 85$, 
  $\shields\approx0.07$, $D^+\approx 12-14$}{%
$\Reb\approx 7600-8900$, $\dratio=2.5$, $\hmean/D\approx 39$,
$\shields\approx0.056 - 0.075$, $D^+\approx 12-14$, $Ga\approx 50$}) 
which is not far from our parameter point. 
Although the experimental data exhibits 
relatively large scatter, it can be seen that the trend of the
DNS data falls within the experimental data cloud.

\revisionold{%
  Finally, let us turn to the phase speed dispersion relation. }
{%
  Finally, let us turn to the celerity of the individual Fourier
  modes of the bedforms. }
The phase angle spectrum $\varphi_j$ of the sediment bed height fluctuation
is defined as
\begin{equation}\label{eq:phase-angle}
 \varphi_j(t) 
      = {\rm atan2}(\Re(\hat{h}_{bj}(t)),\Im(\hat{h}_{bj}(t)))
        \quad \forall j  =0,...,N_x/2
\end{equation}
where ``atan2'' is the four-quadrant inverse tangent.
The instantaneous value
of the phase speed $c_j$ is derived from \eqref{eq:phase-angle} as:
\begin{equation}\label{eq:phase-speed}
c_j(t) = -\frac{1}{\kappa_j}\frac{{\rm d}\varphi_j}{{\rm d}t}(t)\;.
\end{equation}
\revisionold{%
  In the steady dune propagation interval of cases \caseB, \caseC, 
  \caseD\ and \caseE, 
  we have evaluated an average phase speed of the individual modes 
  from their corresponding instantaneous value to infer the mean phase 
  speed dispersion relation $\langle c_j \rangle$. 
  Figure~\ref{fig:propagation-speed-vs-wavelength}\textit{b}
  presents $\langle c_j \rangle$ as a function of the wavelength 
  $\lambda_j$. In all the considered cases, the dispersion relation
  features two distinct types of modes: dispersive and
  non-dispersive. } 
{%
  In the steady ripple propagation interval of cases \caseB, \caseC, 
  \caseD\ and \caseE, 
  we have evaluated an average mode-wise phase speed $\langle c_j
  \rangle$ which is presented in
  figure~\ref{fig:propagation-speed-vs-wavelength}\textit{b} as a
  function of the wavelength  $\lambda_j$. In all the considered
  cases, the celerity data features two distinct types of modes:
  dispersive and non-dispersive. } 
All harmonics with a wavenumber value  smaller than a threshold 
are observed to be non-dispersive, propagating at the corresponding 
mean \revision{dune}{ripple} migration velocity.
To within the statistical uncertainty of the data, the threshold value
is 
\revisionold{%
  fairly the same }
{%
  the same }
in all cases and turns out to be 
$\lambda_j\approx \hmean$.
Note that the non-dispersive modes
are those dominant ones which exhibit a power law relationship
in figure~\ref{fig:dune-geometry-vs-wavelength}.
On the other hand, the remaining harmonics with a wavelength smaller
than the threshold are observed to be dispersive, exhibiting 
approximately the same variation of the phase speed
as a function of the wavelength in all cases. 
The trend is seen to be well defined
when scaled by the friction velocity, increasing from its
value at the threshold location and attaining values of 
$\langle c_j\rangle^+\approx 2$
at wavelengths of approximately $\lambda_j\approx 10D$.
At first glance, the observations of dispersive modes seems incompatible
with the fact that the \revision{dunes}{ripples} are migrating at a constant velocity
maintaining their two-dimensional shape. Looking at
figure~\ref{fig:dune-geometry-vs-wavelength} however, the contribution
of the fast moving modes to the mean \revision{dune}{ripple} shape (and mean migration velocity)
is orders of magnitude smaller than that of the most dominant one, 
thus negligible. 
These modes are footprints of the particle erosion/deposition 
cycle which takes place 
on the top of the \revision{dunes}{ripples}.
Based on field measurements, \citet{Nikora1997b} similarly report that 
fully developed \revision{dunes}{ripples} are composed of non-dispersive modes with wavelength
$\lambda \gg \hmean$, as well as dispersive modes with 
$\lambda \ll \hmean$, corroborating our findings.

\subsection{Mean shear stress at the fluid-bed interface}
\label{subsec:bottom-friction}

For a given imposed flow rate, an increase of the mean interface shear 
stress  with increasing \revision{dune}{ripple} dimensions  is expected since the \revision{dunes}{ripples}
are effectively roughness elements of the sediment bed
\citep[cf.][for instance]{Jimenez2004}. 
This in turn is expected to influence the 
mean rate of particle transport in a given channel, since it is
directly proportional to the mean shear stress. 
The inter-dependency among the evolving \revision{dunes}{ripple}, the 
mean interface shear stress and the mean particle flow rate 
can be assessed by evaluating these quantities 
on the steady interval of each case
(cf.\ table~\ref{tab:numerical-parameters}).

\begin{figure}
   \centering
        \begin{minipage}{3ex}
          \rotatebox{90}
          {\hspace{6ex} $(y-\yref)/\hmean$}
        \end{minipage}
        \begin{minipage}{.45\linewidth}
          \centerline{$(a)$}
          \includegraphics[width=\linewidth]
          {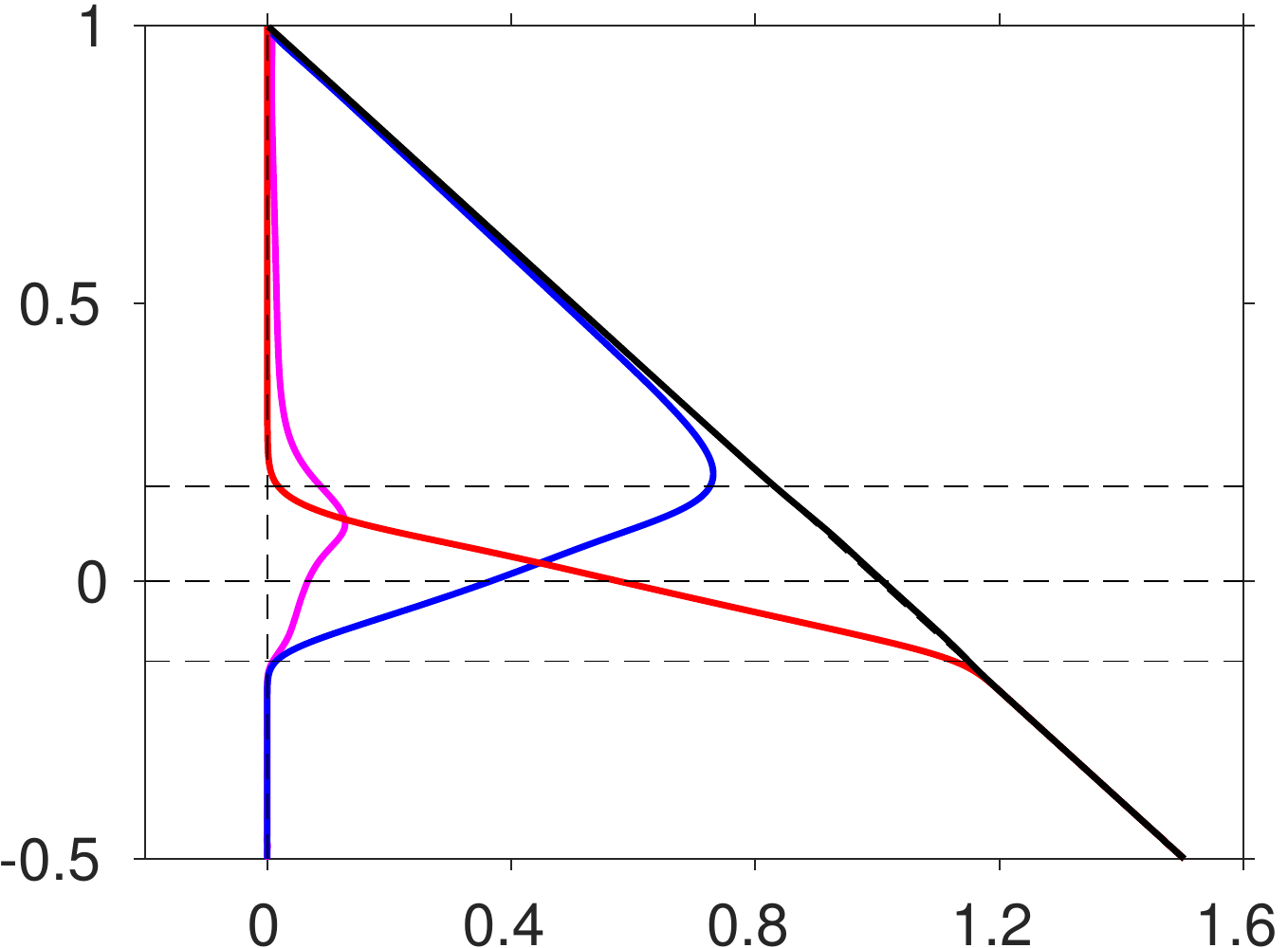}
         \centerline{$\sheartotmean/\rhof \utau^2$}
        \end{minipage}
        \hfill
        \begin{minipage}{3ex}
          \rotatebox{90}
          {\hspace{6ex}$\Ret$}
        \end{minipage}
        \begin{minipage}{.45\linewidth}
          \centerline{$(b)$}
          \includegraphics[width=\linewidth]
          {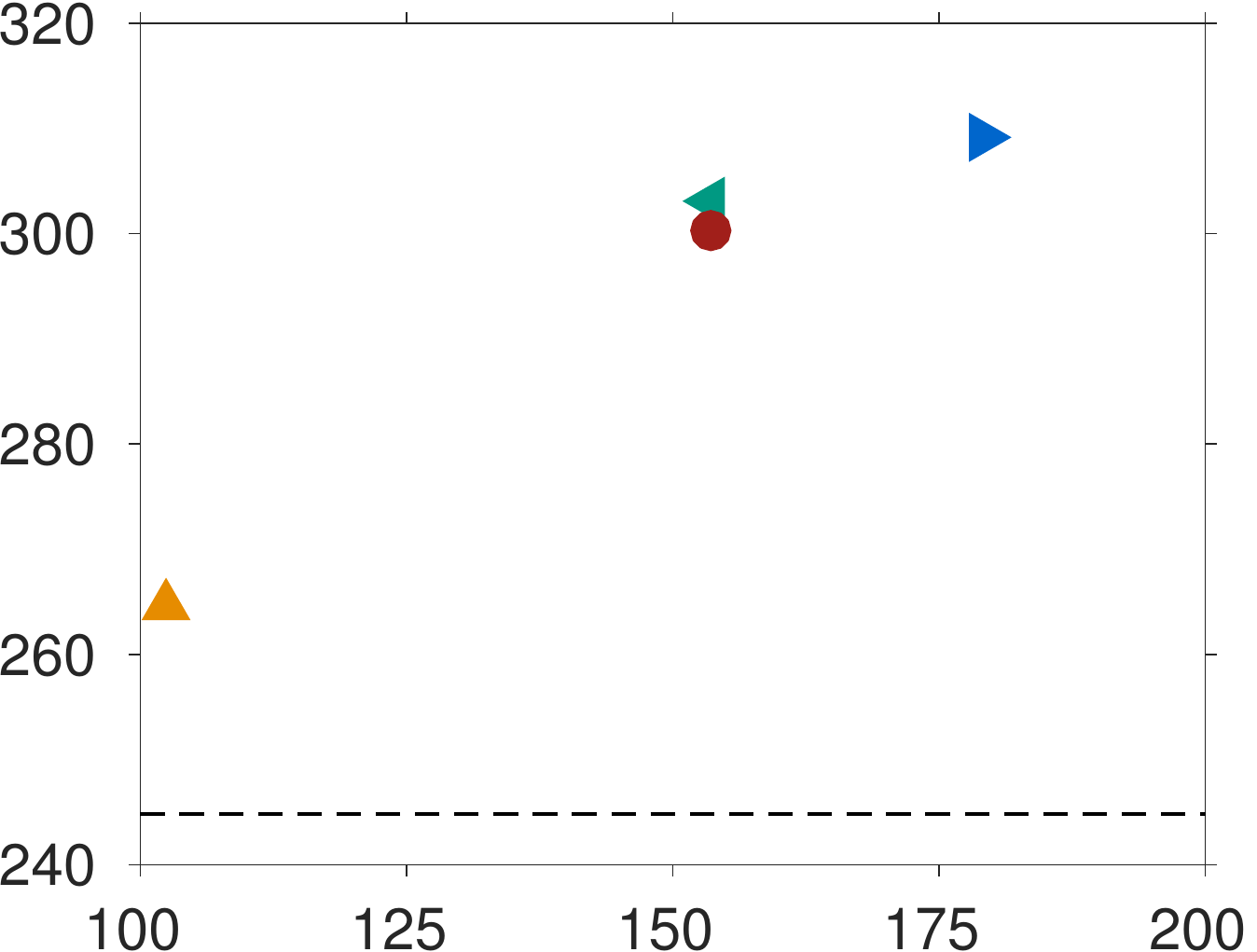}
         \centerline{$\meanlambda/D$}
        \end{minipage}
        \caption{
         (\textit{a})
         Wall-normal profiles of the  different contributions to the 
          total shear stress of case \caseD\
          \revision{}{(computed in the final steady ripple evolution 
          interval as shown in table~\ref{tab:numerical-parameters})}. 
        {\color{magenta}\asolidthick}, $\rhof\nu\partial_y \langle u \rangle$;
        {\color{blue}\asolidthick}, $-\rhof \langle u^\prime v^\prime \rangle$;      
        {\color{red}\asolidthick}, $\int_y^{\Ly}\langle f_x \rangle {\rm d}y$;
         \asolidthick, \sheartotmean.
         The horizontal dashed-lines represent the wall-normal 
         location of mean fluid-bed interface as well as its all-time 
          maximum and minimum extents over the considered observation interval.
        (\textit{b}) Friction velocity based Reynolds number as a function of 
         the mean wavelength $\meanlambda$ 
         \revision{}{in the correspoding steady ripple
         evolution interval}. 
          {\color{myorange}\asolidtriup},~case~\caseB;
         {\color{mygreen}\asolidtrileft}, case~\caseC;
         {\color{myblue}\asolidtriright}, case~\caseD;
         {\color{myred}\asolidcircle}, case~\caseE.
        The horizontal dashed line 
        corresponds to the value of \Ret\ in the featureless case \caseA.
                }
        \label{fig:total-shear-stress-and-retau-vs-wavelength}
\end{figure}
Let us recall that, 
the driving volume force exerted by the imposed mean pressure gradient 
\dpdxmeanline\ in a  
particle-laden channel flow, is balanced by the sum of resisting force
of the fluid shear  stress and stress contribution from the
fluid-particle  interaction. In the context of
the immersed boundary method, the volume force, which
imposes the no-slip condition at the fluid-particle interface, corresponds
to the stress imposed on the system as a result of the fluid-solid
interaction \citep[see e.g.][]{Uhlmann2008,Kidanemariam2013}.
When the momentum equation is averaged over the entire domain 
comprising the fluid and particles, the streamwise momentum
balance reduces to
\begin{equation}\label{eq:total-shear-stress}
       -\dpdxmean (\Ly-y) = %
      \underbrace{
       \rhof\nu\partial_y \langle u \rangle
       - \rhof\langle u^\prime v^\prime \rangle + 
       \int_y^{\Ly}\langle f_x \rangle {\rm d}y
       }_{\sheartotmean}\;
\end{equation}
where the last term in the RHS represents the fluid-solid interaction,
$\langle u \rangle$ is the mean composite velocity, and 
$\langle u^\prime v^\prime \rangle$ is the covariance with respect to
$\langle u \rangle$.
When equation (\ref{eq:total-shear-stress}) is applied to a statistically 
one-dimensional channel flow configuration (for instance case \caseA),
$\langle u^\prime v^\prime\rangle$ represents the momentum flux due to 
turbulent fluctuations. On the other hand, the mean flow in those 
\revision{dune}{ripple}-featuring cases is two-dimensional and $\langle u^\prime v^\prime \rangle$
represents not only the turbulent fluctuations, but also contribution
from the deviation of the mean flow streamlines from the streamwise 
direction
\citep[see e.g.][]{Yalin1977,Nikora2007}.
A detailed analysis of the flow field over \revision{dunes}{bedforms} 
is not part of the 
current study and will appear in a followup paper. Here, in order to 
define a mean interface shear stress,  evaluating equation 
(\ref{eq:total-shear-stress}) suffices. 
In the steady \revision{dune}{ripple} propagation interval of the current configuration
\revision{}{(cf.\ table~\ref{tab:numerical-parameters})},  
\sheartotmean\ should vary linearly as a function of wall-normal
distance. This is confirmed in figure 
\ref{fig:total-shear-stress-and-retau-vs-wavelength}\textit{a}, 
which shows the wall-normal variation of the different contributions 
to the total shear stress for case \caseD. The remaining cases exihibit
similar trend (plots not shown).
The figure highlights that, in the region above the highest crest of
the \revision{dunes}{ripples}, the driving force is dominantly balanced
by the fluid stress term while in the domain below the lowest trough of the 
\revision{dunes}{ripples}, the particle-related forcing entirely balances the 
pressure gradient forcing. 
The mean interface shear stress ($\rhof \utau^2$) 
is thus defined as the value of \sheartotmean\ at the location of the 
mean fluid-bed interface $\yref = \hbedmean$.
It should be noted that, 
in the \revision{dune}{ripple}-featuring cases, due to the roughness introduced 
as a result of the evolving \revision{dunes}{bedforms}, 
a horizontal fluid-bed interface 
does not exist. Nevertheless, a virtual wall could be defined to be 
located at a wall-normal location $\yref = \hbedmean$. 

Figure~\ref{fig:total-shear-stress-and-retau-vs-wavelength}\textit{b}, 
shows the mean interface shear stress (expressed in terms of the 
friction velocity based Reynolds number) as a function of the mean 
\revision{dune}{ripple}
wavelength for the different cases
\revision{}{(computed in the final steady ripple evolution 
          interval as shown in table~\ref{tab:numerical-parameters})}. 
It can be seen that the value of the friction Reynolds number in 
the featureless case \caseA\ is $\Ret \approx 245$. This value is
larger than the value of \Ret\ in a rough-wall channel flow at a 
comparable bulk Reynolds number and at roughly the same particle
Reynolds number \citep[cf.][]{Chan-Braun2011}.
The main difference between the latter case and the present case
\caseA\ is the particle arrangement and particle mobility,
i.e.\ the erosion, entrainment and re-deposition cycle.
Furthermore, when comparing the friction Reynolds number of cases
\caseB, \caseC, \caseD\ and \caseE, it can be observed that there is 
a monotonic increase with increasing wavelength, attaining values 
$\Ret\approx 265$ for $\meanlambda \approx 102D$,
$\Ret\approx 300$ for $\meanlambda \approx 154D$ and
$\Ret\approx 310$ for $\meanlambda \approx 180D$. 
Recalling the fact that the imposed fluid flow rate and the mean fluid
height are essentially the same in all cases, the increase in the shear 
stress among the \revision{dune}{ripple}-featuring cases is entirely a consequence of the 
increase of the amplitude of  the evolving sediment patterns
(i.e.\ the macroscopic roughness height). 
\subsection{Mean particle flow rate}
\label{subsec:particle-flow-rate}
\begin{figure}
  \centering
  \begin{minipage}{3ex}
    \rotatebox{90}
    {\hspace{6ex}$\partflowrate/\qrefi$}
  \end{minipage}
  \begin{minipage}{.45\linewidth}
    \centerline{$(a)$}
    \includegraphics[width=\linewidth]
    {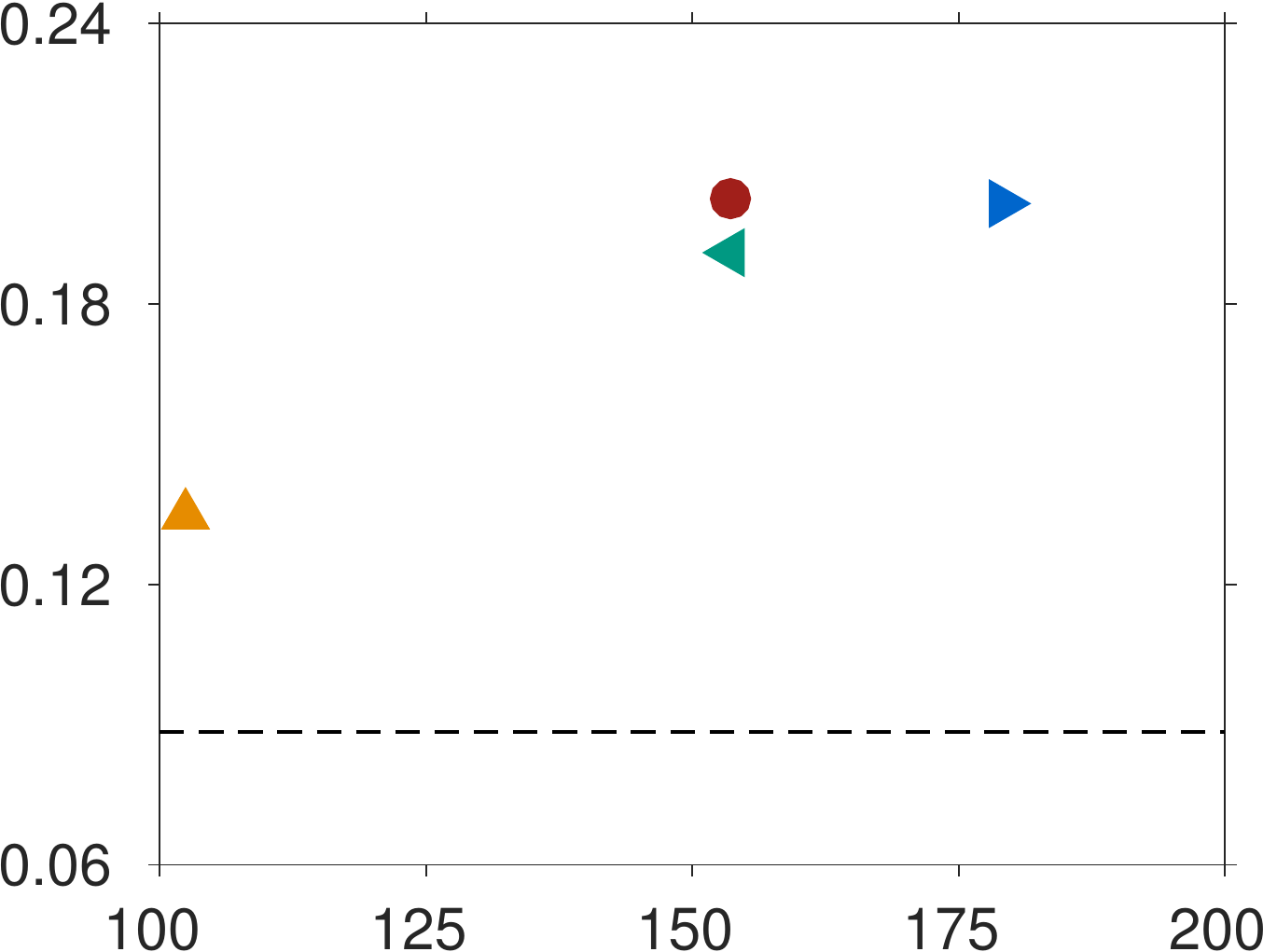}
    \centerline{$\meanlambda/D$}
  \end{minipage}\%
  \hfill
  \begin{minipage}{2ex}
    \rotatebox{90}
    {\hspace{6ex}$\partflowrate_i/\qrefi$}
  \end{minipage}
  \begin{minipage}{.45\linewidth}
    \centerline{$(b)$}
    \includegraphics[width=\linewidth]
    {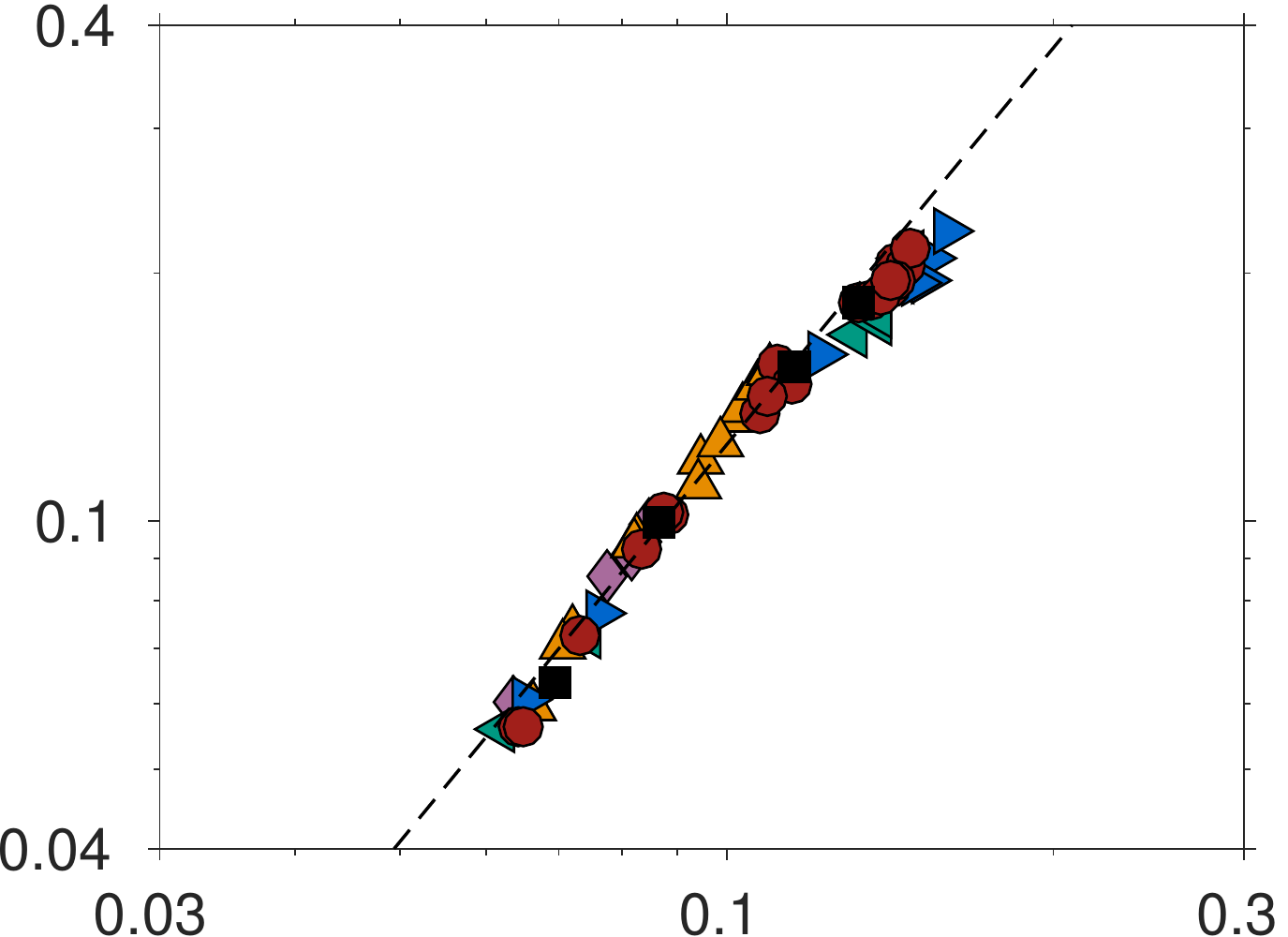}
      \centerline{$\tilde{\theta}$}
  \end{minipage}
  \caption{
    (\textit{a}) Volumetric particle flowrate 
    \partflowrate\ as a function of the mean wavelength
    $\meanlambda$
    \revision{}{(computed in the final steady ripple evolution 
      interval as shown in table~\ref{tab:numerical-parameters})}. 
    {\color{myorange}\asolidtriup},~case~\caseB;
    {\color{mygreen}\asolidtrileft}, case~\caseC;
    {\color{myblue}\asolidtriright}, case~\caseD;
    {\color{myred}\asolidcircle}, case~\caseE.
    The horizontal dashed line corresponds to the value of 
    \partflowrate\ in the featureless
    case \caseA.
    (\textit{b})  
    \revisionold{%
      \partflowrate\ as a function of the excess Shields number
      $\shields-\shieldscrit$.
      The dashed line is the \citet{Wong2006} version
      of the \citet{Meyer-Peter1948} formula for turbulent flow.
    }
    {%
      Particle flow rate $\partflowrate_i$ as a function of the excess
      Shields number $\tilde{\theta}=\shields - \shieldscrit$. 
      The entire observation
      interval of each case is decomposed into smaller intervals of
      approximately 25 bulk time units. 
      {\color{mypurple}\asolidtriup},~case~\caseA;
         {\color{myorange}\asolidtriup},~case~\caseB;
         {\color{mygreen}\asolidtrileft}, case~\caseC;
         {\color{myblue}\asolidtriright}, case~\caseD;
         {\color{myred}\asolidcircle}, case~\caseE;
         {\color{black}\asolidsquare}, case~\caseF.
      A mean particle flow rate and
      bottom shear stress is then computed for each interval
      (thus the notations $\partflowrate_i$ and $\shields_i$).
       Note that values for every fifth interval are plotted for clarity.
      The value $\shieldscrit=0.034$ is used in order to compute
      $\tilde{\theta}$ for the current DNS data, according to the
      empirical law proposed by \cite{soulsby:97}. 
      The dashed line is the \citet{Wong2006} version
      of the \citet{Meyer-Peter1948} formula for turbulent flow which
      reads: 
      $\partflowrate/\qrefi=4.93\,\tilde{\theta}^{1.6}$. 
    }
  }
  \label{fig:particle-flowrate-vs-theta}
\end{figure}

The instantaneous volumetric flow rate of the particle phase
(per unit span), $q_p$, is given by the sum (over all particles) 
of the streamwise particle velocity times the particle volume, divided
by the product of the streamwise and spanwise extent of the domain
\citep{Kidanemariam2014b}, viz
\begin{equation}\label{eq:particle-flow-rate}
q_p(t)  = \frac{\pi D^3}{6 \Lx\Lz}
             \sum_{l=1}^{N_p}u_p^{(l)}(t)\,,
\end{equation}
where $u_p^{(l)}(t)$ is the streamwise component of the instantaneous 
velocity of the $l$th \revision{}{mobile}
particle at time $t$. Averaging $q_p$ over
a given stationary interval results in the mean particle flow rate 
\partflowrate\ in that interval. 
We remark that, for the parameter point considered in the present
study, particles are
dominantly transported as `bedload' material. Only a very small fraction
of the mobile particles is suspended and entrained by the mean flow.
Thus, the particle flow rate defined in 
(\ref{eq:particle-flow-rate}) represents overwhelmingly the former mode of transport.
In figure~\ref{fig:particle-flowrate-vs-theta}\textit{a},
as a consequence of the increased mean bottom friction, the mean particle
flow rate is observed to increase with increasing \revision{dune}{ripple} dimensions.

Of particular relevance to engineering applications is to express
the particle flowrate as a function of the Shields number and to pursue
scaling laws which relate the two quantities. In order to
analyze the particle flowrate obtained in the present work in light
of such scaling laws, the non-dimensional particle
flowrate (normalized by the inertial scale 
$q_{ref} = \ugrav D$) is shown 
\revision{%
  as a function of the excess Shields number in
  figure~\ref{fig:particle-flowrate-vs-theta}\textit{b}. }
{%
  as a function of the excess Shields number 
  $\tilde{\theta}=\shields - \shieldscrit$ 
  in figure~\ref{fig:particle-flowrate-vs-theta}\textit{b}, 
  using the value $\shieldscrit=0.034$ according to the empirical law 
  proposed by \cite{soulsby:97}. } 
Since all the \revision{dune}{ripple}-featuring cases have started from an initially flat sediment
bed, the values of both the shear stress and the particle flowrate in these
cases increase with time from their initial values to their final values at 
the end of the simulation interval.
In order to capture this time evolution, the entire observation interval
of each case is decomposed into smaller intervals of approximately 25
bulk time units. A mean particle flow rate and bottom shear stress is then
computed for each interval, substantially increasing the data samples in
the figure. The duration of the time intervals is chosen
to be much smaller than the time scales of the ripple evolution.
Figure~\ref{fig:particle-flowrate-vs-theta}\textit{b} also shows
the empirical power law of \citet{Wong2006}, which in turn is a modified 
version of the \citet{Meyer-Peter1948} formula for turbulent flows,
\revision{%
  \begin{equation}\label{eq:flwpatt-wong-parker-equation}
    \qpmeanzxt/q_i = A(\shields - \shieldscrit)^\alpha\;,
  \end{equation} 
  where $A=4.93$, $\alpha=1.6$ and $\shieldscrit=0.047$. }
{%
  \begin{equation}\label{eq:flwpatt-wong-parker-equation}
    \partflowrate/\qrefi = A\,\tilde{\theta}^\alpha\;,
  \end{equation} 
  where $A=4.93$ and $\alpha=1.6$. }
Note that Wong \& Parker's formula is valid for a macroscopically flat
sediment bed.  
As is observed in the figure, the DNS data points, which represent both 
plane sediment beds as well as pattern-featuring beds, are in very
good agreement with the scaling law (\ref{eq:flwpatt-wong-parker-equation}).
Although 
the evolution of the ripples does increase the net particle transport rate,
the net bottom shear stress simultaneously increases. As a 
net result, the formation of patterns
does not strongly affect the particle transport scaling as a function of 
the excess shear stress (in the considered observation interval). 

%% file: conclusion.tex
\section{Conclusion}
\label{sec:conclusion}
We have performed several direct numerical simulations of the development of
bedforms over an erodible sediment bed in an open channel flow configuration.
All the simulations were carried out at a parameter point which is 
identical to our previous study \citep{Kidanemariam2014}. 
\revision{%
  The simulations 
  differ only in the adopted streamwise length of the computational domain.
  The latter was systematically varied to address important aspects of 
  pattern formation. }
{%
  The simulations differ only in the adopted streamwise length of the
  computational domain, which was systematically varied in order to
  address important aspects of pattern formation. 
  Ensemble averaging was performed for two select cases in order to
  account for the statistical variability during the transients. 
}
By reducing the domain size, we were able
to find the lower bound of the unstable pattern wavelength, below 
which pattern formation is effectively 
hindered and the sediment bed remains  stable.
The strategy is similar to the  minimal flow unit of \citet{Jimenez1991}. 
For the considered parameter point, it turns out that
a computational box with a streamwise dimension  
 $\Lx~\lesssim~75D$, where $D$ is the particle diameter, 
was not sufficient to 
 accommodate any of the unstable modes, and no sediment features were observed.
 On the contrary, a box with $\Lx~\gtrsim~100D$ accommodated at least one unstable 
 mode. This observation indicates that the cutoff length
 lies in the range 
$75$--$100D$ ($3$--$4$ times the clear fluid height).

Furthermore, 
the influence of the computational domain size on the selection and evolution of
the initial wavelength was assessed by performing one large-scale simulation 
with streamwise box length $L_x\approx 1200D$ ($48\hmean$) and with
approximately $1.1$ million  resolved particles representing the
mobile bed. 
This allowed the determination of the most amplified 
wavelength(s) to be determined with sufficient accuracy.
It turns out that, 
during the initial bed instability, the box was able to  accommodate a number
of ripple units with a mean wavelength 
$\criticallambda \approx 100$--$110D$ (values  which
correspond to the wavelength of the eleventh and twelfth resolved harmonics 
in that case). Based on the
comparison of the selected mean wavelength among the different
simulated cases, it can be concluded that a computational domain
length  which is few integer multiples of $\criticallambda$, due to the
sparsity of the resolved discrete harmonics, severely constrains the
natural ripple initiation and evolution mechanisms.
The domain length adopted by \citet{Kidanemariam2014}, which
accommodated approximately three initial ripple units, was observed to
be marginally sufficient in determining the initial wavelength, while
too small to capture the subsequent evolution.

Based on the analysis of the r.m.s.\ sediment bed height fluctuation,
two regimes of pattern evolution were identified. An initial
short-lived exponential growth regime 
(with a duration of approximately
200 bulk time units) and a subsequent 
non-linear regime. The evolution of the r.m.s.\ bed fluctuation
is observed to be independent of the chosen domain size in the initial
exponential regime, while it exhibited a strong influence of the domain
size in its later stages.
The exponential growth of the sediment bed fluctuation is further 
scrutinized by analyzing the dispersion relation of the
unstable modes, i.e.\  
the growth rate of the individual Fourier harmonics which make up the 
resolved spectrum. The latter, which is difficult to access
experimentally, is an important quantity of interest 
when it comes to assessing the 
validity of the various theoretical models \citep{Charru2013a}. 
It turns out that, 
to within the statistical uncertainty of our data, the modes which are 
initially growing at a substantial rate are those with a 
wavelength approximately in the range $\lambda_j = 100$--$200D$
and are roughly bounded by an upper limit of $\lambda_j = 300$. 

Furthermore, the conditions of our simulations have allowed us to
impose a steady \revision{dune}{ripple} evolution at a desired wavelength. That is, in
the simulation cases  
in which $L_x/\lambda_c = \mathcal{O}(1)$, after the initial
exponential growth interval, the system chooses the maximum possible
mean wavelength $\meanlambda = L_x$ relatively quickly.  Subsequently, the
accommodated ripple evolves steadily maintaining a statistically
time-invariant shape and migration velocity. This allows to address
aspects such as characterization of the two-dimensional asymmetrical
\revision{dune}{ripple} shape, \revision{dune}{ripple} migration velocity and the relation of the latter to
the mean particle flow rate. It was found that the spectrum of the 
sediment bed elevation follows a power-law decay over the first few dominant
modes, with an exponent which exhibited slight dependence on the
evolution of the mean wavelength.
The value of the exponent obtained 
from the DNS data is not too far from the value of ``$-3$'' for 
fully developed bedforms as proposed in the literature
\citep{Hino1968,Nikora1997b}.
Additionally, the relation of the ripple migration velocity 
to the mean particle flow rate was addressed by computing the former 
from the shift of the space-time correlation function of the
sediment bed height as well as from the sediment mass
balance (integration the Exner equation). Both approaches gave 
comparable results highlighting the fact that bedforms migrate as a 
result of erosion of sediment grains from their upstream face and 
deposition at their downstream fronts.

The volumetric particle flow rate of all the simulated cases is 
found to be reasonably well predicted by the  empirical power 
law of \citet{Wong2006}.
It should be noted that the evolution of the patterns has indeed
increased the net particle transport rate, but at the same time,
it has increased the net interface shear stress.
Therefore, it can be concluded that the scaling law proposed by these
authors still holds in the presence of ripples with an amplitude of
the order of a few particle diameters. 
Whether this statement continues to be true at larger pattern
amplitudes needs to be re-assessed by further studies. 

Another aspect of the present problem which requires further attention
is the influence of the evolving sedimentary patterns upon the
turbulent flow. One particularly important question in this context
concerns the shear stress distribution at the fluid/sediment-bed
interface. 
\revision{%
  The present numerical data-set is well-suited for
  evaluating this quantity as well as other details of the flow field
  and the particle motion. }
{%
  The present numerical data-set is well-suited for
  evaluating this quantity as well as other details of the flow field
  and the particle motion (such as the local, instantaneous 
  relation between the particle flux and the shear stress). 
}
This aspect of the problem will be addressed
  in a future contribution. 

%% file: acknowledgments.tex
This work was supported by the German Research Foundation (DFG) through
grant UH242/2-1.
Part of the simulations have been carried out on SuperMUC at the
Leibniz Supercomputing Center (LRZ) of the Bavarian Academy of 
Science and Humanities.
The simulations were also partly performed on the computational 
resource ForHLR I/II of the Steinbuch Centre for Computing (SCC) 
funded by the Ministry of Science, 
Research and the Arts Baden-W\"urttemberg and DFG.
The computer resources, technical expertise and assistance provided by
the staff at these computer centers are thankfully acknowledged.

%% file: appendix.tex
\section{A note on a programming error in the simulations 
         of \citet{Kidanemariam2014}}
\label{sec:appendix-old-vs-new-data}
\begin{figure}[t]
   \centering
        \begin{minipage}{2ex}
          \rotatebox{90}
          {\small \hspace{6ex}$\rmsamplitude/D$}
        \end{minipage}
        \begin{minipage}{.45\linewidth}
          \centerline{(\textit{a})}
          \includegraphics[width=\linewidth]
          {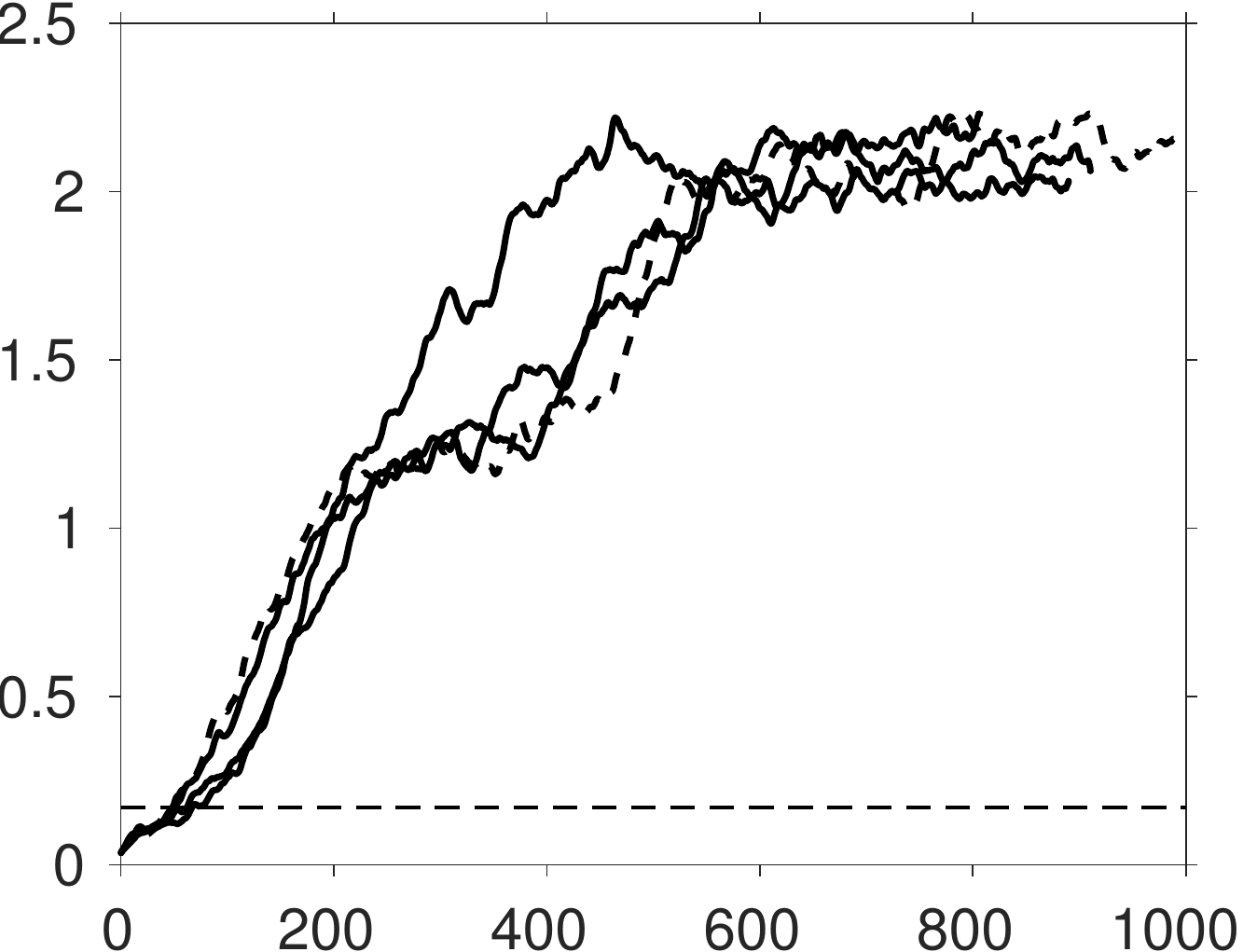}
          \centerline{\small  $t/\tbulk$}
        \end{minipage}
        \hfill
        \begin{minipage}{2ex}
          \rotatebox{90}
          {\small \hspace{6ex}$\meanlambda/D$}
        \end{minipage}
        \begin{minipage}{.45\linewidth}
          \centerline{(\textit{b})}
          \includegraphics[width=\linewidth]
          {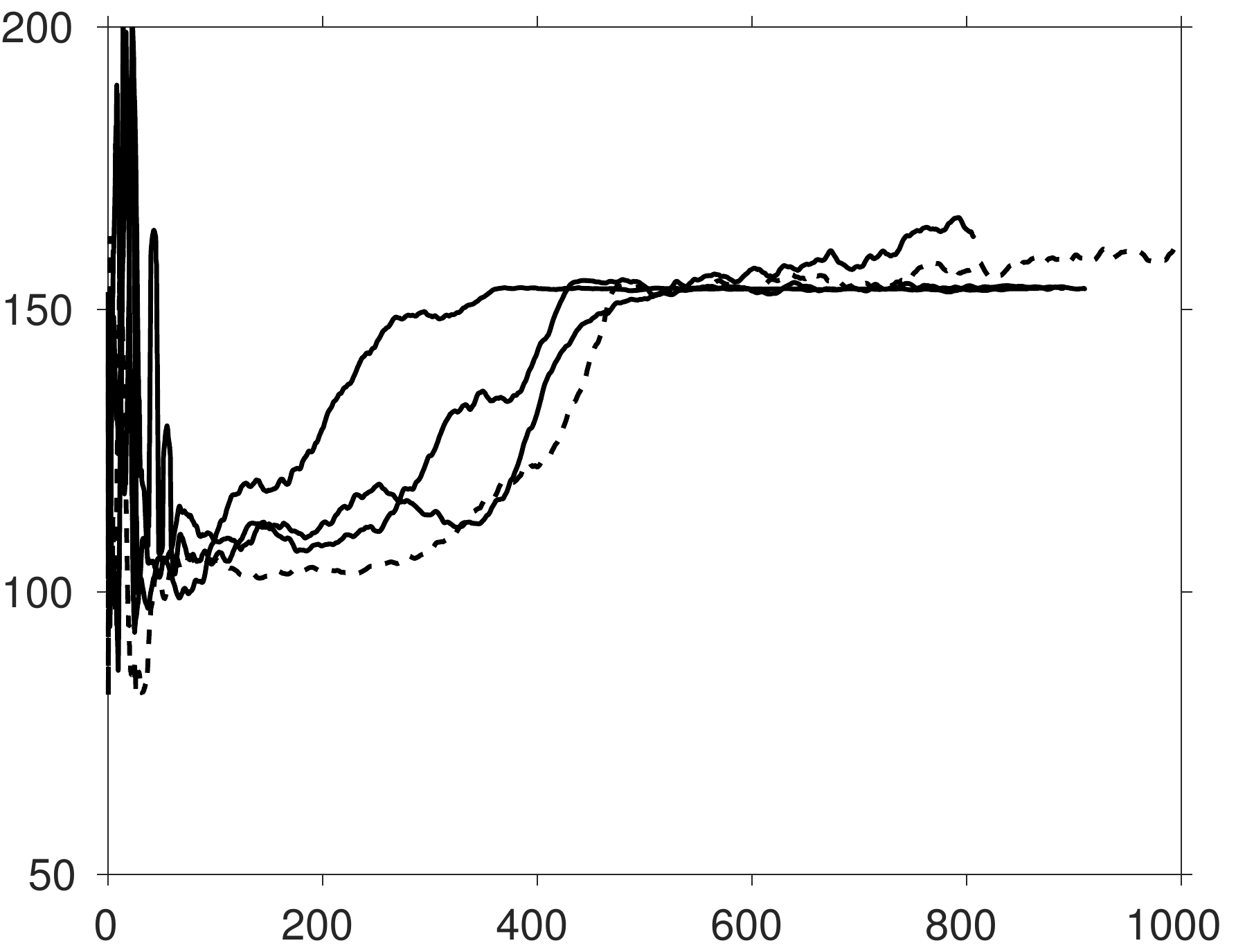}
          \centerline{\small  $t/\tbulk$}
        \end{minipage}
        \caption{%
         Time evolution of (\textit{a}) 
         the r.m.s.\ sediment bed height fluctuation
         and (\textit{b}) the mean wavelength \meanlambda. 
       Solid line corresponds to the different realizations of case 
       \caseE\ while the dashed line represents the data corresponding to
       case `T01' in figure 6 of \citet{Kidanemariam2014}.}      
         \label{fig:before-after-bug-comparison} 
\end{figure}

During the preparation of the present manuscript, we have discovered a
minor programming error in  a subroutine of our simulation code 
which computes the inter-particle collision forces/torques. 
This bug was active during the simulations of our previous contribution 
\citep{Kidanemariam2014}. 
All the DNS data in the present work is generated after correcting 
the aforementioned error. In the following, we 
assess the influence of the bug by comparing the
dune-related quantities extracted from the current data with that reported 
in \citet{Kidanemariam2014}.

Figure~\ref{fig:before-after-bug-comparison} shows the time evolution
of the ripple amplitude and mean wavelength of the three different
simulations of case \caseE\ (which differ from one another 
only in the slightly different initial conditions). 
Additionally, the figure shows the corresponding data from
our previous contribution 
\citep[case `T01' in figure 6 of][]{Kidanemariam2014}.  
It can be seen that the difference between the data with and without 
the bug is within the scatter due to the three independent 
realizations of case \caseE. 
We have further scrutinized the influence of the bug by comparing 
other less-sensitive quantities such as the particle flow rate and the 
mean fluid and particle velocities. The results of these quantities are 
practically not affected by the bug. Thus it can be concluded that 
the impact of the  programming error on the results and conclusions 
made in \citet{Kidanemariam2014} is insignificant.